\documentclass[12pt]{article}    

\newcommand{\be}{\begin{equation}}
\newcommand{\ee}{\end{equation}}
\newcommand{\bea}{\begin{eqnarray}}
\newcommand{\eea}{\end{eqnarray}}

\newcommand{\gs}{\ensuremath{g_s}} 
\newcommand{\ap}{\ensuremath{\alpha'}} 
\newcommand{\ls}{\ensuremath{l_s}} 

\def\p{\partial}

\newcommand{\tr}{\mathop{\rm Tr}}

\def\expec#1{\langle #1 \rangle}

\newcommand{\Rea}{{\mathrm{Re}}}

\newcommand{\cL}{\mathcal{L}}

\newcommand{\cN}{{\mathcal{N}}}
\newcommand{\cO}{{\mathcal{O}}}
\newcommand{\cR}{{\mathcal{R}}}

\newcommand{\bR}{{\mathbf{R}}}
\newcommand{\bS}{{\mathbf{S}}}

\newcommand{\bZ}{{\mathbf{Z}}}

\newcommand{\g}{g_{YM}}
\newcommand{\vy}{\vec{y}}
\newcommand{\hy}{\hat{y}}

\newcommand{\vk}{\vec{k}}
\newcommand{\bm}{{\mathbf{m}}} 
\newcommand{\bl}{{\mathbf{l}}}

\newcommand{\ha}{\widehat{a}}

\newcommand{\hb}{\widehat{b}}
\usepackage{latexsym}  

\begin{document}

\begin{titlepage}

\begin{flushright}
CINVESTAV-FIS-03/08 \\
ICN-UNAM-03/06 \\
hep-th/0305257
\end{flushright}

\vspace{1cm}

\begin{center}
{\huge\bf Conifold Holography}
\end{center}
\vspace{5mm}

\begin{center}

{\large Xavier Amador,$^{\scriptstyle *}$ Elena
C\'aceres,$^{\scriptstyle *}$ \\ \vspace{3mm} Hugo
Garc\'{\i}a-Compe\'an,$^{\scriptstyle *}$
and Alberto G\"uijosa$^{\scriptstyle \dagger}$} \\

\vspace{5mm}

$^{*}$ Departamento de F\'{\i}sica, Centro de Investigaci\'on y de
Estudios Avanzados del IPN, Apdo. Postal 14-740, 07000, M\'exico,
D.F.

\vspace{3mm}

$^{\dagger}$ Departamento de F\'{\i}sica de Altas Energ\'{\i}as,
Instituto de Ciencias Nucleares Universidad Nacional Aut\'onoma de
M\'exico, Apdo. Postal 70-543, M\'exico, D.F. 04510 \vspace{5mm}

{\tt
xamador, caceres, compean@fis.cinvestav.mx, \\
alberto@nuclecu.unam.mx }

\end{center}

\vspace{5mm}

\begin{center}
{\large \bf Abstract}
\end{center}
\noindent We examine the extension of the Klebanov-Witten
gauge/gravity correspondence away from the low-energy conformal
limit, to a duality involving the full, asymptotically Ricci-flat
background describing three-branes on the conifold. After a
discussion of the nature of this duality at the string theory
level (prior to taking any limits), we concentrate on the
intermediate-energy regime where excited string modes are
negligible but the branes are still coupled to the bulk. Building
upon previous work, we are able to characterize the effective
D3-brane worldvolume action in this regime as an IR deformation of
the Klebanov-Witten $\cN=1$ superconformal gauge theory by a
specific dimension-eight operator. In addition, we compute the
two-point functions of the operators dual to all partial waves of
the dilaton on the conifold-three-brane background, and subject
them to various checks.

\vfill
\begin{flushleft}
May 2003
\end{flushleft}
\end{titlepage}
\newpage


\section{Introduction}

The existence of two alternative descriptions of D-brane physics
has been clear ever since Polchinski \cite{polchrr} identified the
localized hyperplanes where open strings can end \cite{dlp} with
the R-R charged black brane solutions of supergravity \cite{hs}.
It seems fair to say, however, that, to date, the precise relation
between these two approaches has not been completely elucidated.
In the early days of the black hole entropy calculations
\cite{entropy}, the understanding was essentially that the two
descriptions were valid in mutually exclusive regimes;
extrapolation from one description to the other would then be
warranted only for protected quantities.  At the same time, the
direct comparison of various quantities in the two pictures, most
notably by Klebanov and collaborators \cite{3/4,klebabs,gkt,gk},
supported the idea that these two perspectives can operate
concurrently, in which case one would be dealing with a
\emph{duality} at the string theory level.

By adopting this second point of view, and considering a
low-energy decoupling limit, Maldacena was able to derive (albeit
heuristically) his celebrated correspondence \cite{malda,gkp,w}.
Given the impressive body of evidence that has accumulated in
support of this gauge/gravity duality \cite{magoo}, one is
compelled to take the starting point of Maldacena's argument
seriously.  It then becomes natural to inquire about the precise
nature and origin of the duality that operates at the level of the
full-fledged string theory, \emph{before} taking any low energy
limits.  A related, but more modest, goal is to study this
correspondence in a regime of energies that are low enough for the
massive string modes to be negligible (thereby cutting the problem
down to a more manageable, essentially field-theoretic, size), but
not low enough for the branes to decouple from the bulk (which
means one is still working away from the Maldacena limit).

The possibility of generalizing in this manner the standard
$\cN=4$ AdS$_5$/SYM$_4$ correspondence away from the conformal
limit has been pursued in various works.  The authors of
\cite{ghkk,dealwis} argued that the duality in question would
equate supergravity on the full asymptotically-flat three-brane
background to the effective theory describing the low-energy
worldvolume dynamics of D3-branes at strong 't~Hooft coupling.
Gubser and Hashimoto \cite{gh} and Intriligator \cite{i} then
characterized the latter theory as an IR deformation of the
$\cN=4$ fixed point by a specific dimension-eight operator (see
also \cite{cm}).

To directly examine this non-renormalizable gauge theory would
clearly constitute a difficult challenge.  The authors of
\cite{dgks} suggested a different line of attack: employing the
conjectural duality to extract information about the dual theory
directly from the three-brane background.\footnote{This approach
is similar in spirit to that of \cite{abks,ms,mr}.} To this end, a
recipe to compute correlation functions was developed, and shown
to satisfy several non-trivial consistency
checks.\footnote{Interestingly, an identical recipe was derived
simultaneously in \cite{dg}, in the context of the supergravity
background dual to NCYM \cite{hi,mr}.} An element that played an
important role in the analysis of \cite{dgks} is the observation
that on the `worldvolume' side of the duality one also needs to
retain the massless closed string modes living in the bulk of
ten-dimensional flat space.  Subsequent tests of this conjectural
intermediate-energy duality have been carried out in
\cite{rvr,ejp}. A different approach to duality in the full
asymptotically-flat three-brane background has been pursued in
\cite{hashimoto} (see also \cite{costa,costa2}).\footnote{Although
not directly related to these developments, for completeness we
would also like to draw attention to the recent proposal
\cite{dbs} for a holographic description of flat space.}

The main purpose of the present paper is to show how this entire
story can be generalized to an $\cN=1$ setting.  Specifically, we
will examine the case where the D3-branes live not in Minkowski
space, but on the conifold \cite{co}, and explore the possibility
of elevating the Klebanov-Witten duality \cite{kw} away from the
conformal limit. We will start out by elaborating on the general
form of the duality at the level of string theory in Section
\ref{stringsubsec}.  This will be followed up in Section
\ref{intersubsec} with a review of the duality in the
intermediate-energy regime that is our main focus. After that, we
will explain in Section \ref{ghsec} how the arguments of
\cite{gh,i} can be carried over to the conifold setting, allowing
us to identify the specific deformation of the Klebanov-Witten
gauge theory that is relevant for the duality in question. Section
\ref{couplsec} describes the way in which the dilaton field
couples to the D-branes, and in particular reviews the form of the
operators in the worldvolume theory that are dual to the dilaton
partial waves. Finally, in Section \ref{dgkssec} we employ the
prescription of \cite{dgks,dg} to work out the two-point
correlators of these operators. A summary of our conclusions is
given in Section \ref{conclsec}.


\section{Dual Descriptions of D3-branes}
\label{dualsec}

\subsection{Worldvolume/geometry string duality}
\label{stringsubsec}

Consider a collection of a large number, $N$, of D3-branes in Type
IIB string theory, placed at radial position $r=0$ in a
ten-dimensional space $M^{3,1}\times Y$ with metric
\be \label{flatmetric}
ds^{2}=\left(-dt^{2}+dx_1^{2}+dx_2^{2}+dx_3^{2}\right)
        +\left(dr^{2}+r^{2} ds_X^{2}\right)~,
\ee
where the coordinates inside the first (second) parentheses denote
directions parallel (transverse) to the branes.  In order for the
six-dimensional transverse manifold $Y$ to be Ricci-flat, we
demand that its five-dimensional base $X$ be Einstein, with
positive curvature.  The discussion in this section will apply to
any such $X$; but the specific examples we have in mind are
$X=\bS^{5}$, in which case $Y$ is simply $\bR^{6}$ (relevant to
the discussion in \cite{klebabs,ghkk,gh,i,dgks,rvr,ejp}), and $X=
T^{11}\equiv [SU(2)\times SU(2)]/U(1)$, in which case $Y$ is the
conifold \cite{co,kw} (relevant to the analysis in the remaining
sections of this paper).  Notice that the point $r=0$ is singular
in all cases except $X=\bS^{5}$.

The physics of this D3-brane system can be described from two
quite distinct perspectives.  On the one hand, we can adopt the
above `worldvolume' perspective, and consider the branes as
hyperplanes with intrinsic dynamics, localized at $r=0$.  The
excitations of the system are then of two kinds: open strings with
endpoints anchored on the branes, and closed strings moving about
in the ten-dimensional metric (\ref{flatmetric}).  On the other
hand, we can subscribe to a `geometry' perspective, replacing the
explicit hyperplanes with the solitonic solution of the string
equations of motion that carries the same charge and mass.  To
leading order in $\ap$ this is of course the black three-brane
solution,\footnote{See \cite{dhss} for a recent analysis of a
particular next-to-leading-order correction to this background.}
with metric \cite{hs,gubsereinstein}
\bea \label{d3metric}
ds^{2} &=& H^{-1/2} \left(-dt^{2} +dx_1^{2} + dx_2^{2} +dx_3^{2}
\right)
+   H^{1/2} (dr^{2} +r^{2} ds_X^{2})~,  \\
H(r)&=& 1+{R^{4}\over r^{4}}~, \qquad R^{4}={4\pi^{4} N\gs\ls^{4}
\over \mbox{Vol}(X)}~, \nonumber
\eea
a constant dilaton $e^{\bar{\phi}}=\gs~$, and $N$ units of
Ramond-Ramond flux through $X$.  The above metric describes a
geometry with an asymptotically flat region $r>R$, and a throat
extending from $r\sim R$ down to a horizon at $r=0$.  On this
background, closed strings are the only allowed excitations.

As explained in the Introduction, following Maldacena's argument
\cite{malda} back to its starting point, we interpret the success
of the AdS/CFT correspondence as (circumstantial) evidence for a
\emph{duality} between the above two perspectives.  To try to make
this `worldvolume/geometry' correspondence more precise, let us
focus attention on a specific physical process: closed string
scattering off the D3-branes.  The concrete claim is then that we
can obtain the same result for the scattering amplitude in two
different ways:
\begin{enumerate}
\item[(\emph{i})] Summing over worldsheets with an arbitrary
number of holes and a fixed number $h$ of handles, with the
Ricci-flat background metric (\ref{flatmetric}).
\item[(\emph{ii})] Computing on a single worldsheet with $h$
handles and no holes, with the black three-brane background
(\ref{d3metric}).
\end{enumerate}
For the lowest $t$-channel poles, and for $h=0$, the agreement
between these two calculations was first verified to leading order
in the number of holes in \cite{kt,ghkm,gm}, and was then argued
to hold to next-to-leading order in \cite{bl}. The expectation for
the agreement to extend to all orders is ultimately a reflection
of open/closed string duality
\cite{dkps,kv,polyakov,dvlmp}.\footnote{Notice that if we work at
unrestricted energies, then to reproduce the \emph{massive}
$t$-channel poles/cuts that are obtained in (\emph{i}), we would
additionally need to consider in (\emph{ii}) non-trivial
backgrounds for the corresponding closed string modes. These
backgrounds would decay exponentially with radial distance, with a
decay constant of order the string length, and so would only be
relevant very close to $r=0$.}

In description (\emph{i}), the worldvolume picture, one naturally
wonders where the curved geometry is hidden.  The answer is that
it is implicit in the sum over holes: each boundary gives rise to
a tadpole for the graviton, as well as for the other closed string
modes.  The claim is then that summing over these tadpoles will
effectively reproduce the non-trivial background.

Conversely, in description (\emph{ii}), the geometry picture, one
appears to be missing the open string modes, which are known from
the worldvolume picture to transform in the adjoint of the
$U(N)\simeq U(1)\times SU(N)$ gauge group.  To be more precise,
what appears to be missing is the $SU(N)$ part: the $U(1)$ degrees
of freedom describe the center-of-mass motion of the D-brane
stack, and must therefore correspond to the zero-mode fluctuations
about the black three-brane background (\ref{d3metric}), i.e., the
Goldstone modes associated with the symmetries broken by the
branes (see, e.g., \cite{acgnr}).  The remarkable lesson of the
AdS/CFT correspondence is that the $SU(N)$ part is in fact also
present in description (\emph{ii}), encoded in the closed string
degrees of freedom that live in the near-horizon region of the
black three-brane background.

Notice that this last point implies that the specific scattering
process we are considering is more generic than one might have
thought at first sight, because through a judicious choice of the
ingoing and outgoing closed string states we can selectively
excite the various degrees of freedom intrinsic to the branes.
Indeed, this use of closed string scattering as a probe of brane
structure lies at the heart of the GKPW prescription \cite{gkp,w}
for computing correlation functions in the standard AdS/CFT
setting, and is also the organizing principle behind the recipe
formulated in \cite{dgks}.

Of course, the main obstacle on the way to making the above
worldvolume/geometry duality more explicit is the different
regimes of validity of descriptions (\emph{i}) and (\emph{ii}). We
know that for $\gs N<1$  the sum in (\emph{i}) is well-defined as
an asymptotic expansion.\footnote{For the discussion of regimes of
validity in this paragraph and the next, in the spirit of
\cite{hp} we are roughly dividing the $\gs N$ half-axis into two
regions, $\gs N<1$ and $\gs N>1$, but of course, strictly speaking
we need to exclude the intermediate region $\gs N\sim 1$ (and,
evidently, the corresponding descriptions are better justified for
$\gs N\ll 1$ or $\gs N\gg 1$, respectively).} In this regime we
can also carry out the computation in (\emph{ii}) (where we have
$R<\ls$), as long as we restrict ourselves to the $r>R$ region
(i.e., we only consider processes in which the momentum
transferred to the D-branes is small in $1/R$ units). But even
without this restriction, notice that the difficulty here is
simply that one cannot in general carry out a \emph{perturbative}
expansion of the non-linear sigma model in powers of $\ap$,
because $R<\ls$.\footnote{Of course, there is the additional
complication of how to treat the background RR field; but this is
clearly just a technical issue. For progress on this question,
see, e.g, \cite{mt,bl,berkovits}.} This difficulty
notwithstanding, the important point for the purpose of
formulating the duality is that sensible meaning can still be
ascribed to description (\emph{ii}), in terms of the
strongly-coupled two-dimensional theory.\footnote{The latter might
even admit a more explicit treatment in terms of a `string bit'
formalism, as in \cite{bo,dmw} (see also \cite{ckky,v,vv} and
references therein).}

In the opposite direction, when $\gs N>1$ we know description
(\emph{ii}) is well-defined; but the \emph{perturbative} sum in
(\emph{i}) is not.  Nonetheless, assuming that there exists a
non-perturbative definition of string theory, we can regard the
sum over holes as a metonym for the corresponding computation in
the strongly-coupled string theory.  Of course, the key question
here is whether this computation can still be meaningfully
formulated in an open string language.  A non-perturbative
formalism that could perhaps be well-suited for this purpose is
open string field theory \cite{osft}. Incidentally, note that in
this latter context it has been argued that closed strings can be
directly expressed in terms of the open string fields (see, e.g.,
\cite{hisft,ag,nadav}), and so do not need to be added as
independent degrees of freedom.  If correct, this would allow a
`purely four-dimensional' formulation of the worldvolume side of
the duality, which would be more in line with what we have gotten
used to in the AdS/CFT setting. For related work, see
\cite{shatash,grsz,gir,est}.

The other piece of string theory lore that might appear to be in
conflict with the worldvolume/geometry duality assumed in
\cite{malda} and advocated here is the Fischler-Susskind mechanism
\cite{lovelace,fs,dr,clny}.  The original application of this
mechanism to worldsheets with holes dates back to the pre-D-brane
era, and was consequently carried out in the context of a
space-filling D$p$-brane (i.e., $p=25,9$ for the bosonic and
super- string cases, respectively).  The issue was that holes were
found to give rise to divergences, and to cancel these it was
necessary to modify the \emph{closed} string background
\cite{dr,clny}.\footnote{As is well-known, in the case of the
unoriented open string, and for a specific choice of the gauge
group, it is possible to cancel the annulus divergence against
similar divergences on the M\"obius strip and the Klein bottle
(or, in modern terms, to adjust the number of D-branes so as to
cancel the charge of the orientifold plane).}

In more detail: the disk, for instance, can be represented as a
sphere with a hole cut out, where one integrates over the radius
and position of the hole.\footnote{These additional moduli
compensate for the fact that the sphere has a larger conformal
Killing group than the disk.} A divergence arises from the lower
limit of the integral over the size of the hole, where the surface
is closing up into a sphere.  In this limit, the leading
contribution to the amplitude can be seen to factorize into three
pieces: the sphere with an additional graviton/dilaton vertex
operator, a graviton/dilaton propagator, and a disk with the
respective graviton/dilaton vertex as its only insertion.  This
last factor is a tadpole amplitude, expressing the possibility for
the space-filling brane to emit (or absorb) a graviton/dilaton.
Since the source is position-independent, the massless particle is
emitted with zero momentum, which means that its propagator must
be evaluated on-shell.  This is then the physical origin of the
divergence.

The insight of Fischler and Susskind is that the theory can be
`renormalized' by shifting the background metric/dilaton, which to
leading order amounts to inserting a single graviton/dilaton
vertex on the sphere, in such a way as to cancel the divergence.
The physical interpretation of this is that, in order to have a
well-defined perturbative expansion, one is forced to perturb
about the vacuum that solves the `loop'-corrected equations of
motion. Shifting to this vacuum in effect amounts to summing up
the tadpoles, to the order one is working at.  In the D-brane
case, then, the shift should take us to the corresponding black
brane background. Indeed, it is precisely through the
Fischler-Susskind mechanism that open string backgrounds are able
to contribute to the Einstein equation (i.e., to the beta
functional for the metric) \cite{dr,clny}.

In a first reading, the above story might seem to imply that one
must necessarily consider explicit D-branes (holes on the
worldsheet) \emph{and} the corresponding black brane geometry
(closed string vertex insertions) \emph{at the same time},
therefore invalidating the duality we have been pursuing.  But
assume for a moment that this is correct, and that at the string
theory level there is only one description, involving explicit
D-branes in the background they themselves generate.  The problem
is then that, starting from this \emph{unique} description and
taking a \emph{single} low-energy limit, one cannot possibly
obtain the two sides of the AdS/CFT correspondence.\footnote{To
try to avoid this problem, it has been argued in \cite{park} (see
also \cite{pst}) that the duality at the level of string theory
has on its geometry side the black three-brane background, and on
its worldvolume side, explicit D3-branes living on the \emph{same}
background. We find it difficult to make sense of a duality
statement of this kind, because there is clearly more on one side
than on the other. An additional problem is that, starting with
this purported equivalence, the decoupling limit does not yield
the correct AdS/CFT statement: there is now a redshift factor on
both sides of the duality, and so one retains the full open string
spectrum instead of just the massless modes. Related work by the
same author may be found in \cite{park2}.}

The observation that shows us the way out of this seeming paradox
is that, in the case of \emph{localized} (as opposed to
space-filling) D-branes, the kinematics no longer forces us to
work right on top of the massless particle pole, a point that has
been stressed in \cite{pt} at the level of the disk.  This means
that for D3-branes, in particular, the individual amplitudes in
the sum (\emph{i}) contain no divergence associated with massless
closed string tadpoles,\footnote{Starting at the annulus level,
worldsheets with holes do give rise to divergences, but these are
\emph{open} string effects associated with D-brane recoil
\cite{ptrecoil,fpr}.  Their counterparts in description
(\emph{ii}) are the usual zero-mode divergences present in the
perturbation expansion about a soliton background (see, e.g.,
\cite{cg}).} and there is consequently no need to invoke Fischler
and Susskind.

The physical point here can be better appreciated by referring
back to a field-theoretic analog, considering a massless scalar
field $\varphi(x)$ with a linear external source $\lambda J(x)$
(see, e.g., \cite{polchinski}).  In such setting one has the
option of computing physical quantities \emph{either} by starting
in the $\expec{\varphi(x)}=0$ vacuum and summing up the
$J$-tadpoles to all orders in $\lambda$, \emph{or} by shifting
from the beginning to the $\expec{\varphi(x)}=\varphi_J(x)$
vacuum, where $\varphi_J(x)$ is the solution to the $\varphi$
equation of motion with the source included. The agreement between
these two methods is of course a triviality in the field theory
context, but its analog in the string setting is precisely the
surprising equivalence (\emph{i})$=$(\emph{ii}), which contains in
particular the remarkable AdS/CFT correspondence.

There are two simple points one can make using this analogy.  One
is that the expansion about $\expec{\varphi(x)}=0$ is well-defined
except in the case where the source $J(x)$ is
position-independent, which would be precisely the analog of the
space-filling D-brane.  The other is that, if one carries out the
sum over tadpoles only to finite order in $\lambda$, then one has
in effect shifted to a vacuum that no longer satisfies
$\expec{\varphi(x)}=0$, but is not yet the one where
$\expec{\varphi(x)}=\varphi_J(x)$, a situation which is analogous
to having to discuss explicit holes and curved geometry at the
same time. Based on this analogy, then, the Fischler-Susskind
mechanism--- or more accurately, its generalization to the case
with no divergences (perhaps along the lines of \cite{pt})--- is
then understood to be precisely the device that perturbatively
implements the shift from description (\emph{i}) to description
(\emph{ii}).  For other perspectives on this question, see
\cite{bl,kv}.  It would clearly be desirable to try to work this
out in a more explicit manner, and we hope to return to this set
of issues elsewhere.  In the remainder of the present paper,
however, we will concentrate on analyzing this
worldvolume/geometry duality in a simplified (essentially
field-theoretic) setting.

\subsection{The duality at low and intermediate energies}
\label{intersubsec}

Having discussed the duality at the level of string theory, let us
now review the different ways in which one can reduce it to more
manageable forms.  {}From this point on we work in the $\gs N\gg
1$ regime, where $R\gg\ls$ and one has relatively good control
over description (\emph{ii}), the geometry picture.  To attain the
desired simplicity, we restrict attention to processes with
energies lower than some cutoff $\Lambda$ that is in turn smaller
than the string scale, $\omega<\Lambda< 1/\ls$.  On the
worldvolume side, this means that we need only consider the
massless closed and open string modes; the description is then in
terms of the effective low-energy bulk theory (supergravity plus
higher-derivative corrections) in the Ricci-flat ten-dimensional
space (\ref{flatmetric}), coupled to the corresponding effective
theory on the D-brane worldvolume.  On the geometry side, one
similarly retains only the lowest closed string modes away from
the branes, but in addition, the presence of the redshift factor
seen in (\ref{d3metric}) implies that by moving towards $r=0$ one
can have excitations with larger local proper energies. More
precisely, at radial position $r$ the effective cutoff on proper
energies is $\Lambda H(r)^{1/4}$, so modes with locally-measured
energies of order $1/R$ are present in the region $r\leq r_{R}$,
where (for $\Lambda\leq 1/R$)
\be \label{rR}
r_{R}= \frac{\Lambda R^{2}}{\left[1-(\Lambda
R)^{4}\right]^{1/4}}~.
\ee
Similarly, modes with string-scale proper energies live inside
$r\leq r_{\ls}$, where
\be \label{rls}
r_{\ls}= \frac{\Lambda R\ls}{\left[1-(\Lambda
\ls)^{4}\right]^{1/4}}~.
\ee

Notice that, for $\Lambda<1/R$, the survival of modes with proper
energies larger than $1/R$ is made possible only by the presence
of the branes. We can consequently regard the region $r\leq r_{R}$
as a rough indication of the portion of the geometry
(\ref{d3metric}) that is expected to be dual to the D3-brane
worldvolume theory (in a sense that will be made more precise
below).

Given the large splitting between the `Kaluza-Klein'\footnote{Note
that, strictly speaking, in the general case a Kaluza-Klein
interpretation is not possible.  That is to say, even though one
can of course always expand the ten-dimensional fields in terms of
$X$-harmonics, the coefficients in this expansion cannot be
regarded as fields living on the `remaining' five-dimensional
spacetime unless the geometry factorizes, which is the case only
in the AdS/CFT limit.} and string energy scales, $1/R\ll 1/\ls$,
even in this simplified setting there are (at least) two
qualitatively distinct energy regimes available.  Maximal
simplicity is attained in the low energy range $\Lambda\ll 1/R\ll
1/\ls$, where, to zeroth order in $\Lambda R$, the branes decouple
from the bulk \cite{klebabs,gkt}, the D3-brane worldvolume theory
becomes conformal, and the portion $r\leq r_{R}$ of the geometry
(\ref{d3metric}) reduces to the near-horizon AdS$_{5}\times X$
form.  We are then left with Maldacena's extraordinary
correspondence \cite{malda,magoo}.

It is customary to formulate the above decoupling limit in units
such that $\ls\to 0$ with $R/\ls$ held fixed, in which $\omega$
can be kept arbitrary (i.e., \emph{after} taking the limit one can
remove $\Lambda\to\infty$).  To focus attention on the $r\leq
r_{R}$ region, one then introduces a new radial
coordinate\footnote{As long as $\gs N$ is held fixed, the
rescaling $u\equiv r/R^{2}$ is morally equivalent to the one
originally employed in \cite{malda}, $u\equiv r/\ls^{2}$. On the
other hand, if we allow ourselves the freedom to change $\gs N$,
then the former rescaling is more useful, as we will see
momentarily. See also the discussion in \cite{pp}.} $u\equiv
r/R^{2}$, which is kept fixed in the limit (one then has
$u_R=\Lambda\to\infty$).  The possibility to express the duality
in this manner is based on the fact that, having found an IR fixed
point by moving along the RG flow associated with the duality, one
is of course free to consider the resulting conformal theory at
unrestricted energies.  In either language, if we additionally
choose to work in the limit of infinite 't~Hooft coupling,
$\g^{2}N\sim\gs N\to\infty$, then $r_{\ls}/r_{R}\to 0$
($u_{\ls}\to 0$), and so the description in the geometry picture
is purely in terms of the supergravity modes.

Notice that, in this low-energy regime, (\ref{rR}) simplifies to
\be \label{uvir}
r_R=\Lambda R^{2}\qquad (u_R=\Lambda)~,
\ee
which manifestly shows that the physics associated with lower
energy scales in the worldvolume theory takes place at smaller
values of the radial coordinate in the geometry picture.  We
recognize (\ref{uvir}) as the statement of the well-known UV-IR
connection \cite{susswi,pp}, linearly translating the bulk radial
coordinate $r$ into an energy scale in the field theory.  This
allows us to identify the redshift factor in (\ref{d3metric}) as
the physical basis for this connection (as well as its relation to
RG flow \cite{hrg}), and to interpret (\ref{rR}) as its
generalization to the entire three-brane background. Notice that
the nature of this connection is more subtle than it is sometimes
assumed: rather than a direct one-to-one correspondence between a
given energy scale $\omega$ and a fixed radial position
$r_{\omega}$, we find that processes with a given energy $\omega$
can take place at all $r\leq r_{\omega}$.

Alternatively, one can work away from the $\Lambda R= 0$ limit,
where the branes no longer decouple from the bulk, the effective
D3-brane theory is not conformally invariant, and the region
$r\leq r_{R}$ of (\ref{d3metric}) is not purely AdS$_{5}\times X$
\cite{klebabs,ghkk,dealwis,gh,i}.  A point that cannot be
overemphasized here is that \emph{one should not mistake the lack
of brane-bulk decoupling for the absence of a duality}--- as
explained in the previous subsection, a duality exists even at the
string level, prior to any limits.

In this intermediate-energy regime the cutoff can be dialed across
the entire sub-stringy range $0<\Lambda<1/\ls$, which contains the
`almost-Maldacena' extreme $0<\Lambda\ll 1/R$, but also includes
cases where much higher energies are allowed.  {}From now on we
restrict ourselves to $\Lambda\ll 1/\ls$; this suppresses stringy
corrections in the bulk effective action and allows us to carry
out the analysis within a supergravity framework.  Notice that the
above restriction translates into $\Lambda R\ll (\gs N)^{1/4}$,
which still allows moderately high energies in the sense of
$\Lambda\gg 1/R$.  This is an important observation because it is
the energy scale $1/R$, and not $1/\ls$, that controls the
higher-derivative terms in the D3-brane worldvolume effective
action \cite{ghkk,dealwis,gh,dgks}. In other words, the theory
becomes conformal in the Maldacena limit $\Lambda R\to 0$, but
remains non-conformal as long as $\Lambda R$ is finite, even if we
allow ourselves to send $\Lambda\ls\to 0$.

An important difference with the AdS/CFT case is that this time we
cannot change to units where $\omega$ is completely unrestricted
unless we simultaneously confine ourselves to the limit of
infinite 't~Hooft coupling, in order to be able to send $\ls\to 0$
while holding $R$ fixed.  This is precisely the `double scaling
limit' introduced by Klebanov in the seminal paper \cite{klebabs},
where the existence of an intermediate-energy duality was first
proposed. Klebanov's limit places us at a particularly tractable
corner of the full parameter space where the intermediate-energy
correspondence discussed in the present subsection is defined. For
simplicity we will work mostly in this corner; but even in this
case it should be borne in mind that the duality under
consideration has been obtained by restricting ourselves to the
substringy domain, and therefore comes with the built-in cutoff
$\Lambda$.  This point is crucial in trying to make sense of the
duality statement, because the theories one equates in this
intermediate-energy domain are inevitably non-renormalizable.

The conclusion is then that, for $0<\Lambda R\ll (\gs N)^{1/4}$,
the effective intermediate-energy theory on the D3-brane
worldvolume, \emph{coupled to} supergravity on the Ricci-flat
space (\ref{flatmetric}), is dual to supergravity on the full
three-brane background (\ref{d3metric}), with an energy cutoff
$\Lambda$ enforced on both sides of the
correspondence.\footnote{Roughly speaking, then, the proposal is
that the worldvolume theory reproduces the physics of the $r<r_R$
region of the three-brane background (which includes much more
than the near-horizon AdS$_{5}\times X$ region), and what remains,
on both sides of the correspondence, is simply supergravity on
$M^{3,1}\times Y$. Of course, the lack of decoupling between the
branes and the bulk makes it difficult to be more precise about
this split.} Since $g_{YM}^{2}N\gg 1$, the worldvolume theory is
strongly coupled, and so is not given merely by the Born-Infeld
action \cite{gh}. To have a more explicit duality statement, it is
thus necessary to determine the intermediate-energy effective
action for a large number of D3-branes at strong 't~Hooft
coupling--- undoubtedly a daunting task. Fortunately, as explained
in \cite{gh,i} for the case $X=\bS^{5}$, string-theoretic
information highly constrains the possible form of the required
action.  In Section \ref{ghsec} we will explain how the arguments
of \cite{gh,i} can be generalized to the case $X= T^{11}$.

We should also say a word about the interactions allowed in this
energy regime.  Since Newton's constant $\kappa\sim\gs\ls^{4}$ is
IR irrelevant, to zeroth order in $\Lambda\ls$ (i.e., in the
strict $\gs N\to\infty$ limit) the supergravity theories in both
the worldvolume and the geometry pictures become free.\footnote{We
are grateful to Juan Maldacena for making this point clear to us.}
The only remaining interactions then take place on the D-brane
worldvolume, with coupling strength $\gs N\to\infty$.  On the
geometry side, the net effect of these interactions on the
supergravity fields should be summarized by the boundary
conditions enforced at the horizon $r=0$. As has been emphasized
in \cite{dgks}, even in this limit the branes \emph{do not}
strictly decouple from the bulk: the coupling in question is not
of order $\kappa$ or $\gs$, but of order $R^{4}\sim\gs N\ls^{4}$.
Excitations on the D-branes can repeatedly transmute into
supergravity modes, wander off into the bulk, and then come back
into the worldvolume.  If we worked away from the strict $\gs
N\to\infty$ limit (to leading order in $\Lambda\ls$, say), then
the supergravity theories would become interacting (albeit
extremely weakly coupled), but if desired we could revert to the
free supergravity setup by sending $\gs\to 0$.


\section{Intermediate-energy Worldvolume Effective Action}
\label{ghsec}

As explained in the Introduction, the main purpose of the present
paper is to study the worldvolume/geometry duality in the conifold
setting, at energies low enough to be able to neglect massive
string modes, but still away from the decoupling limit of
\cite{malda}. Recall from Section \ref{dualsec} that on the
worldvolume side of the duality we are placing $N$ D3-branes in
the ten-dimensional space (\ref{flatmetric}), where from now on we
set $X=T^{11}$.  The latter is a five-dimensional space that can
be defined as the quotient of $SU(2)_{L}\times SU(2)_{R}$ by the
diagonal $U(1)$ generated by $\sigma^{L}_{3}+\sigma^{R}_{3}$;
topologically it is $S^{2}\times S^{3}$. With this choice $Y$, the
six-dimensional cone over $X$, is the conifold, a Calabi-Yau
manifold with a conical singularity \cite{co}. The stack of
D3-branes is placed right at the singularity, $r=0$.

The worldvolume theory describing this system at extremely low
energies has been constructed by Klebanov and Witten \cite{kw}
(see also \cite{mp}).  It is a superconformal gauge theory with
${\mathcal N}=1$ supersymmetry, gauge group $SU(N)\times SU(N)$,
and chiral superfields $A_i, B_k, \ \ i,k = 1,2$.  $A_i$
transforms as $({\bf N},\bar {\bf N})$ and $B_k$ as $(\bar {\bf
N}, {\bf N})$ under $SU(N)\times SU(N)$.  In addition, the theory
has a superpotential
\begin{equation}\label{e:wsuperpotential}
W= \frac{\lambda}{2} \epsilon^{ij} \epsilon^{kl} A_iB_kA_jB_l~,
\end{equation}
which as explained in \cite{kw} would be non-renormalizable if
considered as a perturbation of free field theory, but is in fact
exactly marginal when understood as a deformation of the IR fixed
point of the $\lambda=0$ gauge theory.

On the geometry side, we start out with the three-brane background
(\ref{d3metric}), with $X=T^{11}$.  The near-horizon limit of this
spacetime is AdS$_5 \times T^{11}$, and so, following the
arguments of \cite{malda}, one concludes that Type IIB string
theory propagating on this near-horizon geometry is dual to the
above superconformal gauge theory \cite{kw} (for a review of this
and related dualities, see \cite{hko,afhs}). The $SO(4)\simeq
SU(2)\times SU(2)$ part of the isometry group of $T^{11}$ acts on
the chiral superfields in a simple way; one $SU(2)$ acts on the
$A_i$ and the other on the $B_k$. The remaining part of the
$T^{11}$ isometry group, $U(1)_R$, plays the role of $R$-symmetry
of the $\cN=1$ gauge theory. The four matter fields $A_i,B_k$
carry an $R$-symmetry charge $s=1/2$.

Let us now examine the duality at intermediate energies, away from
the $\Lambda R=0$ limit (where $\Lambda$ is the UV cutoff
introduced in Section \ref{intersubsec}).  More concretely, out of
the various regimes discussed in Section \ref{intersubsec}, we
will henceforth focus attention on the Klebanov limit, $\ls\to 0,
g_s N \rightarrow \infty$, with $R \sim (g_sN)^{1/4} l_s$ held
fixed \cite{klebabs}.  On the geometry side this implies that we
will be working with free supergravity on the full
asymptotically-Ricci-flat background (\ref{d3metric}).  On the
worldvolume side, we are left with free supergravity propagating
on the Ricci-flat spacetime (\ref{flatmetric}), \emph{coupled to}
the intermediate-energy effective Lagrangian
$\cL_{\mbox{\scriptsize D3}}$ on the D3-brane worldvolume, a
non-renormalizable theory whose form we wish to determine.  At low
energies $\cL_{\mbox{\scriptsize D3}}$ must reduce to the
$SU(N)\times SU(N)$ superconformal gauge theory described above
(mirroring the fact that the full geometry (\ref{d3metric})
reduces to AdS$_5 \times T^{11}$ as $r\to 0$), and so the theory
we are after can be characterized as an IR deformation of the
Klebanov-Witten theory \cite{kw}.  We can thus write, without loss
of generality,
\be \label{irreldeform}
\cL_{\mbox{\scriptsize D3}}=\cL_{\mbox{\scriptsize SCFT}}
      +\sum h_{\Delta}R^{\Delta-4}\cO_{\Delta}~,
\ee
where the sum runs over non-renormalizable (IR irrelevant)
gauge-invariant operators $\cO_{\Delta}$ with dimension
$\Delta>4$, and $h_{\Delta}$ is a dimensionless coupling.

Now, a crucial point is that the operators $\{\cO_{\Delta}\}$
appearing in (\ref{irreldeform}) must preserve the same symmetries
as the three-brane background (\ref{d3metric}).  This means that
we are only interested in IR irrelevant operators that are Lorentz
scalars, $SU(2)\times SU(2)\times U(1)_R$ invariant, and $\cN=1$
supersymmetric. In the $X=\bS^{5}$ case, two independent arguments
made it possible to zero in on the desired operators \cite{gh,i}.
As we will now see, both of these can be carried over to our
$X=T^{11}$ case.

The first of these arguments is based on the well-known fact that,
in the AdS/CFT setting, a Lorentz- and gauge-invariant operator of
conformal dimension $\Delta$ is mapped in the geometry side into a
scalar mode of mass $m$, with $\Delta$ and $m$ related through
\cite{gkp,w}
\be \label{mdelta}
\Delta=2+\sqrt{4+m^{2}R^{2}}~.
\ee
This implies that, in the strong-'t~Hooft-coupling regime that is
of interest to us, operators dual to excited string modes (with
masses $m\sim 1/\ls$) acquire large anomalous dimensions
$\Delta\sim(\gs N)^{1/4}\to\infty$. Following \cite{gh}, we can
consequently restrict the sum in (\ref{irreldeform}) to run only
over operators dual to supergravity modes.  Analyzing the full
mass spectrum of Type IIB supergravity on AdS$_5\times T^{11}$,
which has been derived in \cite{Ceresole1,Ceresole2} using
harmonic expansion on $T^{11}$ (see also
\cite{gubsereinstein,jrd}), we find that the only mode compatible
with the required symmetries is dual to the \emph{non-chiral}
superfield $Q^{0}\equiv \tr (W^{2} e^V {\bar W}^{2} e^{-V})$,
where $W_{\alpha}$ and $V$ are respectively the field strength
(chiral) and gauge field (vector) superfields. $Q^{0}$ contains
the descendant $\tr(F_1^{4} + F_2^{4})$, with $\Delta =8$.  Since
this operator belongs to a long multiplet, in principle it is not
expected to have protected dimension; but the Klebanov-Witten
duality predicts otherwise. The existence of such \emph{a priori}
unprotected multiplets with rational conformal dimensions is a
peculiarity of this $\cN=1$ AdS/CFT correspondence
\cite{gubsereinstein,Ceresole1,Ceresole2}.

Applying the logic of \cite{gh} to our case we are thus led to
conjecture that the D3-brane worldvolume theory in the
intermediate-energy regime of interest to us is
\be \label{O8deform}
\cL_{\mbox{\scriptsize D3}}=\cL_{\mbox{\scriptsize SCFT}}
     +h_{4}R^{4}\cO_{8}~,
\ee
where
\be \label{O8}
\cO_{8}\equiv\tr(F_1^{4} + F_2^{4})+\dots~,
\ee
the dots denoting the $\cN=1$ supersymmetric completion.  In
principle, $\cO_{8}$ could also include an admixture of a
double-trace operator of the type $\tr(F^{2})\tr(F^{2})$.  Notice,
however, that $\cO_{8}$ should not be primarily double-trace, for
such a deformation has been argued to be dual to a non-local
string theory \cite{abs,wittenmultitr,bss,ss}, which is certainly
not what we have on our geometry side.

Following \cite{i}, a second argument can be given to show that
the deformation in (\ref{irreldeform}) must involve a single
operator of dimension eight, precisely as we have concluded in
(\ref{O8deform}). Indeed, part of the considerations of \cite{i}
are based just on scaling properties of the metric, and can be
readily applied to any background of the form (\ref{d3metric}).
As explained in \cite{i}, when one considers a background of this
type, with $H(r)=h+R^{4}/r^{4}$, a scaling argument shows that the
absorption probability depends on $h$ only through the effective
coupling $h_{eff} \sim h R^{4}/x^{4}$.  This means that, on the
worldvolume side, the unique deformation parameter $h$ must
multiply an operator in the worldvolume Lagrangian whose dimension
is exactly $\Delta=8$.\footnote{To avoid misunderstandings we
should remark that most of the other deductions of \cite{i} rely
on the existence of sixteen supercharges, and thus cannot be
directly carried over to our $\cN=1$ setting.}

To summarize, with the help of \cite{gh,i} we have been able to
postulate the specific form (\ref{O8deform}) for the D3-brane
worldvolume intermediate-energy Lagrangian relevant to the
worldvolume/geometry duality in the Klebanov limit, $\ls\to 0$
with $R$ fixed.  As was emphasized in \cite{dgks} and elaborated
on in Section \ref{intersubsec} of the present paper, D-brane
physics implies that $\cL_{\mbox{\scriptsize D3}}$ \emph{cannot}
by itself be the complete theory dual to supergravity on the full
three-brane background (\ref{d3metric}): since we are working away
from the Maldacena limit, the worldvolume theory remains coupled
to (in our case free) supergravity in the Ricci-flat bulk
(\ref{flatmetric}).\footnote{Of course, if desired one could in
principle integrate out the supergravity degrees of freedom.
Notice, however, that this would give rise to non-local
interactions on the worldvolume.} To linear order in the
supergravity fields, the relevant brane-bulk interaction
Lagrangian simply couples each supergravity mode to its dual
gauge-invariant operator. In the following section we will discuss
this coupling for the specific case of the partial waves of the
dilaton, which will then enable us in Section \ref{dgkssec} to
compute the two-point functions of the corresponding operators.


\section{Coupling of the Dilaton to the D-branes}
\label{couplsec}

To understand how the dilaton field couples to D3-branes placed at
a conifold singularity, it helps to review this first for the
flat-space ($X=\bS^{5}$) case relevant to
\cite{klebabs,ghkk,gh,dgks}. Letting $x^{\mu}$ and $y^i$ denote
Cartesian coordinates respectively parallel and transverse to the
stack of D3-branes, we know that the interaction is of the form
\be
\label{flatint} S_{\mathrm{int}} = \int d^{4} x\,
   \sum_l R^{2l}\left.\p_{i_1\ldots i_l}\phi(x;y )\right|_{y=0}
        \cO^{i_1\ldots i_l}(x)~,
\ee
with $\cO^{i_1\ldots i_l}(x)$ the gauge theory operators
determined explicitly in \cite{KTR}.  The appearance in
(\ref{flatint}) of derivatives with respect to the Cartesian
coordinates $y^i$ correlates with the fact that these operators
have the schematic form $\cO^{i_1\ldots i_l}(x)\sim \tr
\left[(F^{2}+\ldots )\Phi^{i_1}\cdot\cdot\cdot\Phi^{i_l}\right]$,
where $\Phi^{i}= X^{i}/R^2$ are the scalar fields associated with
the transverse position of the branes.

Both in the worldvolume and in the geometry sides of the duality
it is natural to decompose the dilaton field into partial waves,
\be \label{flatpw}
\phi(x;\vy)= \phi(x;r,\hy)= \sum_l \phi^{(l)}_{\bm}(x;r)
Y_{l\bm}(\hy),
\ee
where $Y_{l\bm}(\hy)\equiv C^{i_1\ldots
i_l}_{l\bm}\hy^{i_1}\cdot\cdot\cdot\hy^{i_l}$ is one of the
$(l+3)(l+2)^{2}(l+1)/12$ $\bS^{5}$-harmonics corresponding to the
$l$th partial wave \cite{LMR}, and $\hy\equiv\vy/r$ is a unit
vector.  It is important therefore to understand how the partial
waves $\phi^{(l)} $ are related to the differentiated dilaton
appearing in (\ref{flatint}). Comparing the Taylor expansion of
$\phi(x;\vy)$ about $\vy= 0$ against the right-hand-side of
(\ref{flatpw}), it is easy to deduce that \cite{KTR}
\bea \label{flatphil}
\phi^{(0)}(x;r)&\propto& \left[\phi(x;\vy)+\frac{1}{12}r^{2}
  \p_i\p_i\phi(x;\vy)+\ldots\right]_{\vy= 0}~, \nonumber \\
\phi^{(1)}_{i}(x;r)&\propto&r\left[\p_i\phi(x;\vy)
       +\frac{1}{16}r^{2}\p_i\p_j\p_j\phi(x;\vy)+\ldots\right]_{\vy=
       0}~,
\nonumber\\
\phi^{(2)}_{ij}(x;r)&\propto& r^{2}\left[\p_i\p_j\phi(x;\vy)
  +\ldots\right]_{\vy= 0}~,
\eea
and so on (the dots represent terms with more derivatives).

Now, recall from Section \ref{dualsec} that the explicit stack of
D3-branes lives by definition on the worldvolume side of the
duality, where the spacetime metric is (\ref{flatmetric}). For the
flat space case, and for a given partial wave, the solution to the
radial dilaton equation of motion that is regular at $\vy= 0$ is
$\phi^{(l)}(r)\propto r^{-2}J_{l+2}(qr)$, where
\be \label{q}
q^{2}\equiv -k^{\mu}k_{\mu}~,
\ee
with $k_{\mu}$ ($\mu=0,1,2,3$) the momentum along the D3-brane
worldvolume directions. The above solution vanishes like $r^l$ as
$r\to 0$ (where the D-branes are located), and so the gauge theory
operators must couple not to $\phi^{(l)}(r)$ but to
$r^{-l}\phi^{(l)}(r)$, which is finite at the location of the
branes.  And indeed, using (\ref{flatphil}) we see that this is
what (\ref{flatint}) is telling us. We can consequently rewrite
the coupling in the form
\be
\label{flatint2} S_{\mathrm{int}} = \int d^{4} x\,
   \sum_{l,\bm} \left[\left(\frac{R^2}{r}\right)^{l}
   \phi^{(l)}_{\bm}(x;r)\right]_{r=0}\cO^{l\bm}(x)~,
\ee
which shows the explicit pairing between each of the components of
a given dilaton partial wave and the gauge theory operator with
the same $SU(4)$ quantum numbers.

Let us now move on to the conifold. For the relevant $\cN= 1$
theory \cite{kw}, the superfields $A_i,B_k$ (or more precisely,
their scalar component fields $a_i,b_k$) play a role analogous to
the scalar fields $\Phi^i$ of the $\cN= 4$ case, so it is
convenient to use the corresponding coordinates on the conifold.
To set our notation, let us first review how these coordinates are
introduced \cite{co,kw}. Start with the equation that defines the
conifold,
\be \label{conifold}
z_1^{2}+z_2^{2}+z_3^{2}+z_4^{2}=0,
\ee
and define a radial variable $r$ through
\be \label{r}
r^{3}=|z_1|^{2}+|z_2|^{2}+|z_3|^{2}+|z_4|^{2}.
\ee
Letting
\be
z_1=z'_1+iz'_2,\quad z_2=z'_2+iz'_1,\quad z_3=z'_3-iz'_4, \quad
z_4=z'_4-iz'_3,
\ee
we can rewrite (\ref{conifold}) and (\ref{r}) in the form
\cite{kw}
\be \label{conifold2}
z'_1 z'_2-z'_3 z'_4=0
\ee
and
\be \label{r2}
r^{3}=|z'_1|^{2}+|z'_2|^{2}+|z'_3|^{2}+|z'_4|^{2}.
\ee
Next, introduce coordinates $a_i,b_k$ ($i,k=1,2$) that solve
(\ref{conifold2}), through
\be \label{ab}
z'_1=a_1b_1, \qquad z'_2=a_2b_2, \qquad z'_3=a_1b_2, \qquad
z'_4=a_2b_1,
\ee
in terms of which
\be \label{r3}
r^{3}=(|a_1|^{2}+|a_2|^{2})(|b_1|^{2}+|b_2|^{2}).
\ee
The rescaling $a_i\to \lambda a_i$, $b_k\to\lambda^{-1}b_k$ leaves
(\ref{ab}) invariant, and we can use this freedom to arbitrarily
set
\be \label{R}
|a_1|^{2}+|a_2|^{2}=|b_1|^{2}+|b_2|^{2}\equiv \cR^{2}.
\ee
This describes two three-spheres of radius $\cR=r^{3/4}$ embedded
in $\bR^{4}$, so $T^{11}$ (i.e., the conifold at a fixed $r>0$) is
$\bS^{3}\times\bS^{3}$ modulo the freedom to shift the phase of
$a_i,b_k$, which was not eliminated by the condition
(\ref{R}).\footnote{As explained in \cite{kw}, when determining
the moduli space of the gauge theory one encounters an equivalent
pair of restrictions, and the moduli space is then seen to be (the
product of $N$ copies of) the conifold.}

We can again expand the dilaton field into harmonics
\cite{Ceresole1,Ceresole2},
\be \label{conipw}
\phi(x;r,\ha,\hb)= \sum_{\bl\bm} \phi^{(\bl)}_{\bm}(x;r)
   Y^{\bl\bm}_{(0)}(\ha,\hb)~,
\ee
where $\bl\equiv(l,j,s)$ label a given representation of the
$T^{11}$ isometry group, $SU(2)\times SU(2)\times U(1)_{R}$, and
$\bm\equiv (l_{3},j_{3})$ identify a specific component of the
$(2l+1)(2j+1)$-dimensional representation. The harmonics
$Y^{\bl\bm}_{(0)}$ are eigenfunctions of the scalar Laplacian on
$T^{11}$,
\be \label{harmonics}
\Box Y^{\bl\bm}_{(0)}(\ha,\hb) = -H_0(\bl)
Y^{\bl\bm}_{(0)}(\ha,\hb)~,
\ee
with eigenvalues \cite{gubsereinstein,Ceresole1,Ceresole2}
\be \label{h0}
H_0(l,j,s) = 6\left[l(l+1) +j(j+1) - \frac{s^{2}}{8}\right]~.
\ee
The $SU(2)\times SU(2)$ quantum numbers $(l,j)$ must be both
integers or both half-integers,  and the R-symmetry charge $s$
satisfies\footnote{In general the harmonics carry one more label,
$q\in\bZ$, which refers to their charge under the $U_H(1)$ group
that defines $T^{11}$ as a coset space (and is not to be confused
with (\ref{q})). For the harmonic $Y^{(l,j,s)}_{(q)}$, the
restriction over $s$ reads  $|q + s| \leq 2l,$ $|q - s| \leq 2j,$
$2l_3 = q + s,$ $2j_3 = q - s.$ For scalar harmonics $q=0$, and
this is the only case we will consider in the present work.}
$|s|\leq 2l$, $|s|\leq 2j$.

As we will see in the next section, the solution to the radial
dilaton equation of motion that is regular at $r=0$ (where the
D3-branes are located) is $\phi^{(\bl)}\propto
r^{-2}J_{\upsilon}(qr)$, where
\be \label{upsilon}
\upsilon(l,j,s)= \sqrt{H_{0}(l,j,s)+4}~.
\ee
For  $r\to 0$, then, we have $\phi^{(\bl)}\sim r^{\upsilon-2}$,
from which we conclude that the coupling between the branes and
the bulk dilaton field should take the form
\be
\label{coniint} S_{\mathrm{int}} = \int d^{4} x\,
   \sum_{\bl,\bm} \left[\left(\frac{R^2}{r}\right)^{\upsilon-2}
   \phi^{(\bl)}_{\bm}(x;r)\right]_{r=0} \cO^{\bl\bm}(x)~.
\ee
This is then the conifold analog of the flat-space interaction
(\ref{flatint2}). By construction, the above products of powers of
$r$ and partial waves $\phi^{(\bl)}$ are finite at the location of
the branes, so this is the only form of the coupling that could
make sense.\footnote{The appearance of explicit factors of $r$ in
(\ref{flatint2}) and (\ref{coniint}) is quite analogous to what
happens in the standard GKPW recipe \cite{gkp,w} for, say,
AdS$_5$: in that case the equation of motion for a scalar field
$\varphi$ with mass $m$ has two independent solutions, behaving
like $z^{4-\Delta}$ and $z^{\Delta}$ for small $z$, where $z=
R^{2}/r$ (so that $z=0$ is the boundary of AdS) and $\Delta$ is
given by (\ref{mdelta}). The corresponding gauge theory operator
$\cO$ (of dimension $\Delta$) then couples not to the boundary
value of $\varphi(z)$, but to the boundary value of
$z^{\Delta-4}\varphi(z)$, which is finite at $z=0$.} The factor of
$R^2$ has been inserted in order for $S_{\mathrm{int}}$ to be
dimensionless, knowing that $\cO^{\bl\bm}$ has dimension
$\Delta=2+\upsilon$.\footnote{\label{normfoot}For the time being
we are taking $\phi$ to be conventionally normalized (with a
factor of $1/2\kappa^2$ in front of its kinetic term), and so
dimensionless. In the next section we will work instead with a
\emph{canonically} normalized dilaton. To retain (\ref{coniint})
as it stands, we will then need to absorb a factor of $R^4$ into
$\cO^{\bl\bm}$.}

The operators $\cO^{\bl\bm}(x)$ lying in multiplets with protected
dimensions have been identified in \cite{Ceresole1, Ceresole2}.
Two types of multiplet-shortening conditions are relevant in this
context. The dilaton partial waves with $l=j=s/2$ lie in short
multiplets and are dual to descendant operators of the type
$\tr[F_1^{2}(ab)^s+F_2^{2}(ba)^s+\ldots]$ contained in the chiral
superfields $\Phi^s \equiv \tr[W_1^{2}(AB)^s+W_2^{2}(BA)^s ]$,
where $W_{1\alpha}$ ($W_{2\alpha}$) is the field strength
superfield for the first (second) factor of the $U(N)\times U(N)$
gauge group, and one symmetrizes separately over the indices of
the $A_i$ and the $B_k$. (Waves with $l=j=-s/2$ are then related
to the corresponding antichiral superfields.) The partial waves
with $j=l-1=s/2$ are in `semi-long' multiplets, and are dual to
descendant operators contained in the `semi-conserved' superfields
$I^s\equiv\tr[Ae^{V_2}\bar{A}e^{-V_1}W_1^{2}(AB)^s]$. (Similarly,
waves with $l=j-1=s/2$ are related to
$\tr[Be^{V_1}\bar{B}e^{-V_2}W_2^{2}(BA)^s]$.)

All of the remaining partial waves lie in long multiplets and
couple to operators whose dimension at the conformal fixed point,
$\Delta=2+\upsilon$, is irrational. These are linear combinations
of descendants of the various gauge-invariant superfields that can
be constructed with the basic building blocks $W^{2}(AB)^k$,
$Ae^{V}\bar{A}e^{-V}$, $Be^{V}\bar{B}e^{-V}$, and
$W_{\alpha}e^{V}\bar{W}_{\dot{\alpha}}e^{-V}$. Generally one
expects mixing between all of the operators carrying a given set
of quantum numbers. Notice also that in our setting all of the
above operators would receive additional corrections (e.g., of the
type $\tr[F_1^{4}(ab)^n+F_2^{4}(ba)^n]$) due to our departure from
the conformal limit.


\section{Correlation Functions}
\label{dgkssec}

In this section we will calculate the two-point functions of the
operators dual to all dilaton partial waves in a conifold
background.  In addition to providing information about the
departure of the gauge theory from the conformal fixed point, this
will allow us to test the intermediate-energy prescription of
\cite{dgks} in a background that is non-spherical and
non-maximally-supersymmetric.

The basic idea behind the recipe of \cite{dgks} for computing
correlation functions is to compare the amplitude for dilaton
propagation in the geometry and worldvolume pictures. The
corresponding dilaton propagators, $G$ and $G_0$, are defined as
solutions to the ten-dimensional linearized dilaton equation of
motion,
\be \label{propeq}
\partial_M \left[\sqrt{-g} g^{MN} \partial_N G\right] =
i\delta^{(10)}(x^M-x'^M)~,
\ee
in the respective backgrounds (\ref{d3metric}) and
(\ref{flatmetric}), with $X=T^{11}$. The ten-dimensional
coordinates $x^M$ ($M=0,\ldots,9$) consist of the directions
$x^{\mu}$ ($\mu=0,\ldots,3$) parallel to the branes, the radial
direction $r$ away from the branes, and some convenient set of
$T^{11}$ coordinates--- e.g., the $\ha,\hb$ defined in the
previous section.

It is natural then to solve (\ref{propeq}) by separation of
variables, expressing the dependence on the worldvolume directions
in terms of plane waves $\exp(ik_{\mu}x^{\mu})$, and expanding $G$
and $G_0$ in terms of the scalar $T^{11}$ harmonics as in
(\ref{conipw}),
\be \label{proppw}
G(r,\ha,\hb;r',\ha',\hb')=\sum_{\bl,\bm}
Y^{\bl\bm}_{(0)}(\ha,\hb)G^{(\bl)}(r,r')
Y^{\bl\bm}_{(0)}(\ha',\hb')~.
\ee Using this and the eigenvalue equation (\ref{harmonics}) in
(\ref{propeq}), the radial propagator for the $\bl\equiv(l,j,s)$
dilaton partial wave on the three-brane background
(\ref{d3metric}) can be seen to satisfy
\be \label{d3propeq}
\left( \frac{1}{r^{5}}\partial_r (r^{5}\partial_r) -
\frac{H_0(\bl)}{r^{2}} + q^{2} H(r) \right)
G^{(\bl)}(k^{\mu};r,r') = \frac{i\delta(r-r')}{r^{5}}~,
\ee
where $q$ is given by (\ref{q}). The corresponding propagator on
the Ricci-flat background (\ref{flatmetric}), $G_0^{(\bl)}$,
satisfies the same equation with the replacement $H(r)\to 1$.

As explained in \cite{gh} (see also \cite{mmlz,clvp}), the
homogeneous version of (\ref{propeq}) can be related to Mathieu's
equation. Following the procedure of \cite{dgks}, the propagator
on the geometry side of the correspondence (with boundary
conditions such that the associated flux moves \emph{away} from
the source) can then be worked out to be
\be \label{d3prop}
G^{(\bl)}(k^{\mu};r,r')={\pi \over 4  r^{2}r'^{2}}
   H^{(1)}(\nu,\ln ({r_{>}\over R}))
   \left\{H^{(2)}(\nu,\ln ({r_{<}\over R}) ) +
   \frac{\chi-{1\over\chi}}{\chi-{1\over \eta^{2}\chi}}
   H^{(1)}(\nu,\ln ({r_{<}\over R}) )
   \right\},
\ee
where $r_{<}$ ($r_{>}$) is the smaller (larger) of $r$ and $r'$.
We are adopting here the notation of \cite{gh,dgks}:
$H^{(1,2)}(\nu,z)$ are associated Mathieu functions of the third
and fourth kind, respectively, $\nu$ (not to be confused with the
constant $\upsilon$ introduced in (\ref{upsilon})) is the `Floquet
exponent' defined in \cite{gh,mmlz}, $\eta\equiv\exp(i\pi\nu)$,
and $\chi\equiv\varphi(-\nu/2)/\varphi(\nu/2)$, with
$\varphi(\pm\nu/2)$ two of the coefficients involved in the
definition of Mathieu functions. For further characterization of
$\nu$, $\eta$ and $\chi$ (which are all $\upsilon$-dependent
functions of $qR$), see (\ref{nuchi}) below, references
\cite{gh,mmlz}, and Appendix A of \cite{dgks}.

Again following \cite{dgks}, the propagator in the worldvolume
side of the correspondence can be expressed in terms of Hankel
functions,
\be \label{flatprop}
G^{(\bl)}_{0}(k{^\mu};r,r')={\pi\over 4 r^{2}r'^{2}}
   H^{(1)}_{\upsilon}(qr_{>}) \left\{H^{(2)}_{\upsilon}(qr_{<})
   + H^{(1)}_{\upsilon}(qr_{<})
   \right\}~.
\ee
The linear combination inside the braces, $2J_{\upsilon}$, is
singled out by the requirement that the propagator be regular at
$r=0$. Our conclusion then is that the conifold propagators
 $G^{(\bl)}$ and $G^{(\bl)}_0$ have the same form
as those obtained in \cite{dgks} for the spherically-symmetric
case, with the replacement $l+2\to\upsilon$.

Now that we have determined the dilaton propagators $G^{(\bl)}$
and $G^{(\bl)}_0$, we can proceed to compute the two-point
function of the gauge theory operators dual to the partial waves
of the dilaton,
\be \label{twoptdef}
\Delta^{(\bl)}_2(q^{2}) \equiv
\expec{\cO^{\bl\bm}(k)\cO^{\bl\bm}(-k)}~.
\ee
Notice from the propagator equations (\ref{propeq}) or
(\ref{d3propeq}) that we are working now with a \emph{canonically}
normalized dilaton. As in \cite{dgks}, this implies for the gauge
theory operators an overall normalization $\cO^{\bl\bm}(x)\sim
R^4\tr[F^2\cdot\!\cdot\cdot]$ (see footnote \ref{normfoot}).

The prescription of \cite{dgks} to determine $\Delta^{(\bl)}_2$ is
simply to equate the amplitudes obtained on both sides of the
correspondence for dilaton propagation between two points far away
from the branes. Following \cite{dgks}, and given the form of the
coupling (\ref{coniint}), for the regime under consideration this
can be seen to amount to the statement that, in the limit
$r,r'\to\infty$,
\be \label{recipe}
G^{(\bl)}(r,r')=G^{(\bl)}_{0}(r,r')+
\left[G^{(\bl)}_{0}(r,r'')\left(R^2\over
r''\right)^{\upsilon-2}\right]_{r''=0}\!\!\!\Delta^{(\bl)}_2
\left[\left(R^2\over
r''\right)^{\upsilon-2}G^{(\bl)}_{0}(r'',r')\right]_{r''=0}.
\ee
This in turn implies that
\be \label{delta2}
\Delta^{(\bl)}_{2}= \lim_{r,r'\to\infty\atop r''\to0}
\frac{G^{(\bl)}(r,r')-G^{(\bl)}_{0}(r,r')} {G^{(\bl)}_{0}(r,r'')
(R^2/r'')^{2\upsilon-4}G^{(\bl)}_{0}(r'',r')}~.
\ee

As $r,r'\to\infty$, the Mathieu functions in (\ref{d3prop})
asymptote to Hankel functions, and the propagator on the
three-brane can be shown to simplify to \cite{dgks}
\be \label{d3propasym}
G^{(\bl)}(r,r')\to {1 \over 2 q}\left({1\over r r'}\right)^{5/2}
   \left\{e^{iq(r-r')} +
   \frac{\chi-{1\over\chi}}{\chi-{1\over \eta^{2}\chi}}
   e^{iq(r+r')+i2\theta}
   \right\}~,
\ee
with $\theta=-\pi(2\nu+1)/4$. The Ricci-flat-space propagator
similarly reduces to
\be \label{flatpropasym}
G^{(\bl)}_0(r,r')\to {1 \over 2 q}\left({1\over r r'}\right)^{5/2}
   \left\{e^{iq(r-r')} +
   e^{iq(r+r')+i2\theta_0}
   \right\}~,
\ee
with $\theta_0=-\pi(2\upsilon+1)/4$.

For the denominator of (\ref{delta2}), we also need the asymptotic
form of $G^{(\bl)}_0$ as one of its arguments tends to infinity
and the other to zero,
\be \label{flatprop0infty}
G^{(\bl)}_0(r,r'')\to \sqrt{\pi \over 2q}
\frac{r^{-5/2}r''^{\upsilon-2}}{\Gamma(\upsilon+1)} \left({q\over
2}\right)^{\upsilon}
   e^{i(qr+\theta_0)}~.
\ee

Using (\ref{d3propasym}), (\ref{flatpropasym}) and
(\ref{flatprop0infty}) in (\ref{delta2}), we see that all
dependence on $r,r',r''$ cancels out, which is important in order
for the limit (and consequently $\Delta^{(\bl)}_{2}$) to be
well-defined. Our final result for the two-point function then
follows as
\be \label{delta2exp}
\Delta^{(\bl)}_{2}(q^{2})=-{2^{2\upsilon}\over\pi}
\left[\Gamma(\upsilon+1)\right]^{2}R^{8-4\upsilon}
 q^{-2\upsilon}\left\{e^{i\pi\upsilon}
    \frac{\chi-{1\over\chi}}{\eta\chi-{1\over \eta\chi}} -1\right\}~.
\ee

An important check on the above result is to see whether it
satisfies the optical theorem, i.e., whether it is correctly
related to the absorption probability for the corresponding
dilaton partial wave. As in \cite{gh,dgks}, the latter takes the
form
\be \label{pabs}
 P_{\mathrm{abs}}=
 1-\left|\frac{\chi-{1\over\chi}}{\eta\chi-{1\over\eta\chi}}
\right|^2~.
\ee
Expanding the field operator for the $\bl$-th partial wave
according to
\be \label{aadagger}
\phi^{(\bl)}(x;r)=\int^{\infty}_{0}dq\,\sqrt{q}\int^{\infty}_{-\infty}
\frac{d^{3}\vk}{(2\pi)^{3}2\omega_{\vk,q}}
r^{-2}J_{\upsilon}(qr)\left[e^{ik\cdot x}a^{(\bl)}_{\vk,q}+
e^{-ik\cdot x}a^{(\bl)\dagger}_{\vk,q}\right]~,
\ee
where $ k^{0}=\omega_{\vk,q}\equiv\sqrt{q^{2}+\vk^{2}}$, using the
form (\ref{coniint}) for the dilaton-brane coupling, and following
\cite{dgks}, one can derive the precise statement of the optical
theorem on the worldvolume side of the
correspondence,\footnote{\label{ifoot}Notice that equation (45) in
\cite{dgks} erroneously claimed that it was the imaginary part of
$\Delta_2$ that is relevant for the optical theorem. This error
was however compensated by the fact that an $i$ was incorrectly
omitted from the right-hand side of equations (14) and (15) in
that paper. As a result, the expressions for $G$, $G_0$ and
$\Delta_2$ given there were all missing a factor of $i$.}
\be \label{optical}
P_{\mathrm{abs}} = 2\Rea\, \left(\frac{\pi R^{4\upsilon-8}
q^{2\upsilon}\Delta^{(\bl)}_2}{2^{2\upsilon}[\Gamma(\upsilon+1)]^{2}}
\right) -\left|\frac{\pi R^{4\upsilon-8}
q^{2\upsilon}\Delta^{(\bl)}_2}{2^{2\upsilon}[\Gamma(\upsilon+1)]^{2}}
\right|^2~.
\ee
And indeed, inserting our result (\ref{delta2exp}) for
$\Delta^{(\bl)}_{2}$ into (\ref{optical}), one can verify that the
probability (\ref{pabs}) is correctly reproduced.

The correlators (\ref{delta2exp}) can be expanded in powers of
$S\equiv(qR)^{2}$. To understand the analytic structure of this
expansion, it is convenient to employ the results of
\cite{gh,mmlz} to write
\be \label{nuchi}
\nu(S)=\upsilon+\lambda(S^2), \qquad \chi(S)=e^{-\nu(S)\ln
S}\beta(S^{2})\equiv e^{-\lambda(S)\ln S+\gamma(S)}~,
\ee
where $\lambda$ and $\beta$ are ($\upsilon$-dependent) functions
of $S^{2}$ that are analytic in a neighborhood of $S=0$
($e^{\gamma}$ is then seen to be analytic only for integer
$\upsilon$).

Curiously, the case $\upsilon=2$, which corresponds to the mode
that is symmetric on $T^{11}$ (i.e., $l=j=s=0$), is qualitatively
different from all the rest. For this mode $\lambda\equiv i\mu$
and $\gamma\equiv i\alpha$ turn out to be purely imaginary
(implying that $\eta\equiv\exp(i\pi\nu)$ is real and $\chi$ a pure
phase) \cite{gh,mmlz}. As expected, in this case one finds the
same result as for the spherically-symmetric mode in \cite{dgks},
\be \label{cotformula}
\Delta^{(\mathbf{0})}_2(S) \propto iS^{-2}\sinh(\pi\mu)
    {\mathrm{cot}}\left[-\mu \ln(-S) + \alpha\right]
    +S^{-2}\left(1-\cosh(\pi\mu)\right)~.
\ee
The analytic functions $\mu$ and $\alpha$ have leading behavior
$S^{2}$ and $S^{0}$, respectively, so $\Delta^{(\mathbf{0})}_2$
has a sensible low-energy expansion \cite{dgks},
\bea \label{delta0lowenergy}
i\Delta^{(\mathbf{0})}_2(S) &\propto& S^{2}\ln(-S)
  +S^{4}\left[-\frac{1}{24}\left(\ln(-S)\right)^{2}
+\frac{7}{72}\ln(-S)\right] +S^{6}\left[\frac{17}{6912}
(\ln(-S))^{3} \right.
\nonumber \\
{}&{}&  \left.
 -\frac{161}{18432}(\ln(-S))^{2}+\left(\frac{5\pi^{2}}{13824}
 +\frac{5561}{663552}\right)\ln(-S)\right]+ \cdots~.
\eea

Notice that the leading term in (\ref{delta0lowenergy}) is as
dictated by conformal invariance for an operator of conformal
dimension $\Delta=4$. This confirms that, at low energies, our
result correctly reduces to the expected two-point function in the
Klebanov-Witten gauge theory. This is non-trivial because, as seen
in (\ref{delta2}), we have taken the limit $r,r'\to\infty$, i.e.,
we are probing the three-brane geometry from afar. The second term
in (\ref{delta0lowenergy}) is the leading correction due to our
departure from the conformal fixed point. As expected,
$\Delta^{(\mathbf{0})}_2(S)$ has a branch cut for real positive
$S$, which comes entirely from the first term in
(\ref{cotformula}). As noted already in \cite{dgks}, the presence
of the second term in (\ref{cotformula}) implies that for real
negative $S$ (i.e., away from the cut) the two-point function is
\emph{not} purely imaginary.\footnote{In \cite{dgks} the two-point
function was expected to be purely \emph{real}, due to the missing
factor of $i$ mentioned in the previous footnote.} However, since
the real terms are analytic, one is tempted to simply discard
them.

For all $\upsilon>2$, $\lambda$ and $\gamma$ turn out to be real
(implying that $\eta$ is a pure phase and $\chi$ is real)
\cite{gh,mmlz}. Let us first consider the cases where
$\upsilon\in\bZ$. {}The results of \cite{Ceresole1,Ceresole2}
reviewed at the end of the previous section imply that $\upsilon$
is an integer for \emph{half} of the operators with protected
dimensions. In more detail: for operators in short multiplets, we
have $l=j=|s/2|$, and consequently $\upsilon=3l+2$; for operators
in semi-long multiplets, $l=j-1=|s/2|$ (or $j=l-1=|s/2|$), and
therefore $\upsilon=3l+4$ ($3j+4$). So in both cases
$\upsilon\in\bZ$ only if $l$ and $j$ are integer (rather than
half-integer). Under these circumstances one finds that
\be \label{cothformula}
\Delta^{(\bl)}_2(S) \propto iS^{-\upsilon}\sin(\pi\lambda)
    \coth\left[-\lambda \ln(-S) + \gamma\right]
    +S^{-\upsilon}\left(1-\cos(\pi\lambda)\right)~.
\ee
As in (\ref{cotformula}), we observe a clean separation between
the imaginary and real parts. Notice, however, that the leading
behavior of both terms is now of order $S^{2-\upsilon}$ ($\lambda$
is still of order $S^2$, and $\gamma$ is now of order $\ln(S)$),
implying that $\Delta^{(\bl)}_2$ is \emph{singular} in the IR.
This important point was missed in the analysis of the
spherically-symmetric case in \cite{dgks}, where $\upsilon$
becomes $l+2$ and is therefore always integer. It is difficult to
know what to make of this puzzling behavior. Taken at face value,
the appearance of terms that are singular at low-energy is
reminiscent of results obtained in the context of noncommutative
theories, due to the phenomenon of  UV/IR mixing \cite{mvrs,vrs}.
It would perhaps be worth exploring this possible connection (see
\cite{dgks} for a discussion of additional parallels with
noncommutative theories).

In the general case $\upsilon\notin\bZ$ (which covers the case of
generic operators in long multiplets, as well as operators in
protected multiplets with half-integer $l$ and $j$), rewriting
(\ref{delta2exp}) in terms of the variables $\lambda$ and $\gamma$
defined in (\ref{nuchi}) one finds that $\Delta^{(\bl)}_2(S)$
becomes proportional to
\be \label{generalformula}
S^{-\upsilon}\left\{e^{i\pi\upsilon}\frac{\cos(\pi\lambda)
\sinh\left[-\lambda \ln(-S) + \gamma\right] -i\sin(\pi\lambda)
\cosh\left[-\lambda \ln(-S) + \gamma\right]}{\cos(\pi\upsilon)
\sinh\left[-\lambda \ln(-S) + \gamma\right] +i\sin(\pi\upsilon)
\cosh\left[-\lambda \ln(-S) + \gamma\right]}-1\right\}~.
\ee
The leading low-energy behavior of this expression can again be
shown to be singular, of order $S^{2-\upsilon}$, but the analytic
structure is clearly much more intricate than in the
$\upsilon\in\bZ$ case. Notice in particular that there is now
similar non-analytic behavior in both the imaginary \emph{and} the
real parts of $\Delta^{(\bl)}_2$. By applying the recipe of
\cite{dgks} in the conifold setting, we have thus learned that in
general it would be a mistake to simply discard the unwanted real
terms in $\Delta^{(\bl)}_2$, as one was tempted to do in the
spherically-symmetric situation. In any event, as was emphasized
in \cite{dgks}, it is the entire expression (\ref{delta2exp}) that
is needed in order to satisfy the optical theorem (\ref{optical}).

The expression (\ref{delta2exp}) (or (\ref{generalformula})) can
be expanded in powers of $S$, to find a series of the general form
\be \label{deltaseries}
\Delta^{(\mathbf{\bl})}_2(S)=-{2^{2\upsilon}\over\pi}
\left[\Gamma(\upsilon+1)\right]^{2}R^{8-2\upsilon}
\sum^{\infty}_{n,m=0}
\sum^{m}_{k=0}C_{nmk}(\upsilon)S^{(2n-1)\upsilon+2m}
\left(\log(-S)\right)^{k} + \mathrm{analytic}~.
\ee
As we have just noted, the coefficients $C_{nmk}$ in this
expansion are in general \emph{complex} (see, e.g., the explicit
expressions in (\ref{coefs}) below). The reason for this can be
understood by looking back at (\ref{recipe}). If we had worked in
momentum space for all ten of the spacetime directions, then the
propagators $G^{(\bl)}$ and $G_{0}^{(\bl)}$ would have been as
usual purely imaginary, and so (\ref{recipe}) would imply that
$\Delta^{(\mathbf{\bl})}_2$ is purely imaginary as well. We are
working in momentum space for the directions parallel to the
branes, but in position space for the transverse directions, and
as a result, the propagators (\ref{d3prop}) and (\ref{flatprop})
are manifestly complex. The definition (\ref{recipe}) then implies
the same for $\Delta^{(\mathbf{\bl})}_2$. Notice, however, that
our two-point function does meet the physical requirement of
unitarity, as expressed in the optical theorem (\ref{optical}).
One should also bear in mind that the appearance of complex
coefficients in correlation functions is not unique to the
approach of \cite{dgks}: it is also seen, e.g., in the two-point
functions derived in \cite{ms,mr} (a fact that was not noticed in
those works). For additional discussion of this issue, see
\cite{dgks}.

Notice that (\ref{deltaseries}) reads
\be \label{deltallowenergy}
i\Delta^{(\bl)}_{2}(S)\propto \mbox{IR-singular terms\,}+
            C_{100}S^{\upsilon}
        +C_{111} S^{\upsilon+2} \ln(-S)
     +\mbox{subleading terms}.
\ee
The term of order $S^{\upsilon}$ is of the form expected by
conformal invariance for the two-point function of an operator of
conformal dimension $\Delta=2+\upsilon$ in the Klebanov-Witten
gauge theory, but now due to the IR singularities this is
\emph{not} the dominant term in $\Delta^{(\bl)}_{2}$ at low
energies.  The $S^{\upsilon+2}$ term shown above (as well as a
similar term without the logarithm) has the form required for the
leading correction due to our departure from the conformal fixed
point. For the spherically-symmetric case, it was shown in
\cite{rvr} that the analogous term could be reproduced via an
AdS/CFT three-point function computation. It would be very
interesting to check if the same can be done here, using the
standard GKPW recipe \cite{gkp,w} in the AdS$_5\times T^{11}$
setting to work out the three-point function
$\expec{\cO^{\bl\bm}\cO^{\bl\bm}\cO_8}$, obtained by inserting in
$\Delta^{(\bl)}_{2}$ the dimension-eight operator (\ref{O8})
coming from the deformed worldvolume Lagrangian (\ref{O8deform}).
Evidently more work is needed also to clarify the meaning of the
IR singularities observed in $\Delta^{(\bl)}_{2}$, and in
particular whether they are somehow an artifact of the recipe of
\cite{dgks}.

With these future applications in mind, we record here the first
few of the coefficients in the expansion (\ref{deltaseries}),
worked out for the case $\upsilon\notin\bZ$ using Mathematica:
\bea \label{coefs}
C_{000} &=& 0 \nonumber
\\
C_{010} &=& - \frac{i \pi}{4\upsilon-4 {\upsilon^{3}}} \nonumber
\\
C_{011} &=& 0 \nonumber
\\
C_{020} &=&-\frac{\pi^{2}}{2(4 \upsilon-4 \upsilon^{3})^{2}} +
\frac{i \pi (8-35 \upsilon^{2} + 15 \upsilon^{4})}{64 \upsilon^{3}
(-4+{\upsilon^{2}})(-1+ \upsilon^{2})^{3}} \nonumber
\\
C_{100} &=& - \frac{2^{1-4 \upsilon} \pi^{2}}{\Gamma(\upsilon)^{2}
\Gamma(1+\upsilon)^{2}} + \frac{i
 4^{-2\upsilon} \pi^{2} \csc^{2}(\pi\upsilon) \sin(2\pi\upsilon)}
 {\Gamma(\upsilon)^{2}
\Gamma(1+\upsilon)^{2}} \nonumber
\\
C_{110} &=& \frac{2^{-1-6 \upsilon} \pi^3 \csc(\pi\upsilon)
[4^\upsilon \cos (\pi  \upsilon) \Gamma(\upsilon)
\Gamma(1+\upsilon) + 8 \upsilon (-1+ \upsilon^2) k_2]}{\upsilon
(-1+ \upsilon^2) \Gamma(\upsilon)^3 \Gamma(1+\upsilon)^3}-
 \\
&&\frac{i 2^{-1-6 \upsilon} \pi^3 \cot (\pi\upsilon)\csc(\pi
\upsilon) [4^\upsilon \cos (\pi  \upsilon) \Gamma(\upsilon)
\Gamma(1+\upsilon)+8 \upsilon (-1+\upsilon^2) k_2]}{ \upsilon (-1+
\upsilon^2)\Gamma(\upsilon)^3 \Gamma(1+\upsilon)^3} \nonumber
\\
k_2 &\equiv& \frac{4^\upsilon \cos (\pi \upsilon) \Gamma(\upsilon)
\Gamma(1+\upsilon)}{4 \upsilon-4 \upsilon^{3}} + \frac{1}{\pi
\upsilon^{2} (-1+ \upsilon^{2})^{2}} \times \nonumber
\\
& & \left(4^{-1+\upsilon}\Gamma(\upsilon)\Gamma(1+\upsilon)
\sin(\pi\upsilon) \bigg[1+(\upsilon-\upsilon^{2})[\upsilon (-1+\ln
4)+\ln 4]\right. \nonumber
\\
& & \left.\left.- 2 \upsilon (-1+\upsilon^{2})
\frac{\Gamma'(\upsilon)}{\Gamma (\upsilon)} \right] \right) \nonumber
\\
C_{111}&=& -
 \frac{{4^{1-2 \upsilon}} {{\pi }^2}}
 {(4 \upsilon-4 {\upsilon^{3}})
 {{\Gamma(\upsilon)}^2} {{\Gamma(1+\upsilon)}^2}}+\frac{i
 {2^{1-4 \upsilon}} {{\pi }^2}
 {{\csc^2(\pi\upsilon)}} \sin (2 \pi  \upsilon)}
 {(4 \upsilon-4 {\upsilon^{3}}) {{\Gamma(\upsilon)}^2}
{{\Gamma(1+\upsilon)}^2}} \nonumber
\\
C_{122} &=& -\frac{4^{1-2 \upsilon} \pi^{2} \csc (\pi \upsilon)(-i
\cos (\pi \upsilon) + \sin (\pi\upsilon))}{(4 \upsilon -
4\upsilon^{3})^{2} \Gamma(\upsilon)^2 \Gamma(1+\upsilon)^2}
\nonumber \eea

\section{Conclusions} \label{conclsec}

Building upon previous work \cite{klebabs,ghkk,dealwis,gh,i,dgks},
in this paper we have studied the higher-energy precursor of the
AdS/CFT correspondence, concentrating for the most part on the
$\cN=1$ duality that equates the two alternative descriptions of a
stack of D3-branes living on the conifold \cite{kw}. In Section
\ref{stringsubsec} we have discussed some of the issues that arise
in trying to formulate a precise duality statement at the level of
string theory (prior to taking any limits). Our main point has
been that the difficulties one encounters are not limitations of
principle, but of practice--- not unlike what one runs into in the
various instances of S-duality in string theory, or, to some
extent, in the AdS/CFT case itself.

In Section \ref{intersubsec} we have reviewed the various ways in
which the above `worldvolume/geometry' string duality can be cut
down to more accessible sizes. These include in particular the
Maldacena limit $\ls\to 0$ with $R/\ls$ fixed \cite{malda}, and
the Klebanov limit $\ls\to 0$ with $R$ fixed \cite{klebabs}. An
important point is that, from the outset, the simplification
process necessarily involves the introduction of a cutoff
$\Lambda<1/\ls$ on both sides of the correspondence. Of course, in
the Maldacena case, \emph{after} flowing down to the conformal
fixed point one can choose to remove the cutoff; but in the
general case the presence of the cutoff is crucial in order to
make sense of the duality, because the field theories one equates
are ineluctably non-renormalizable.

Along the way, we have seen in equation (\ref{uvir}) that the
well-known UV-IR connection ultimately arises from the redshift
factor of the three-brane metric (\ref{d3metric}). It would be
interesting to try to relate this discussion to the curious form
of the UV-IR relation deduced for the full three-brane background
(in the flat-space, as opposed to conifold, setting) in
\cite{dgks}. It was shown there that the profile of a one-point
function in the presence of a source localized in the bulk at
first decreases in size as the source moves to larger values of
$r$, but then starts to grow again when the source has moved past
$r=R$.\footnote{The same type of behavior was obtained in the
context of the supergravity dual of NCYM in \cite{drnc}, using a
recipe for correlation functions \cite{dg} that is identical to
that of \cite{dgks}.} In terms of the discussion of Section
\ref{intersubsec} it is in a sense natural for the asymptotically
(Ricci-)flat region $r\to\infty$ (where the massless supergravity
modes are always present) to correspond to low energies, but it
would be worth trying to further elucidate the full picture.

As explained in Section \ref{intersubsec}, as long as $\gs N\gg 1$
there is a wide intermediate-energy region $0<\Lambda R<(\gs
N)^{1/4}$ where one stays away from the AdS/CFT endpoint $\Lambda
R=0$ (which implies in particular that the branes remain coupled
to the bulk), and yet one is still able to neglect--- modulo
redshift effects--- the massive string modes (which means that the
duality can be formulated in essentially field-theoretic terms).
{}From Section \ref{ghsec} on we chose to restrict our analysis to
the Klebanov limit \cite{klebabs}, $\ls\to 0$ with $R$ fixed
(implying $\gs N\to\infty$), which places us at a particularly
tractable corner of the full parameter space available in this
intermediate-energy region.

In this particular limit, one confronts a duality that equates
free supergravity on the conifold-three-brane background
(\ref{d3metric}) with the intermediate-energy effective action on
the D3-brane worldvolume, \emph{coupled to} free supergravity in
the ten-dimensional Ricci-flat geometry (\ref{flatmetric})
\cite{klebabs,dealwis,ghkk,dgks}. In Section \ref{ghsec}, by
adapting the symmetry arguments of \cite{gh,i} to our setting, and
employing the results of \cite{Ceresole1,Ceresole2} for the
AdS$_{5}\times T^{11}$ spectrum, we have been able to conjecture
the specific form (\ref{O8deform}) for the required worldvolume
action.

In Section \ref{couplsec} we have reviewed the way in which the
various dilaton partial waves $\phi^{(l,j,s)}$ couple at linear
order to the worldvolume action, as well as the form of the gauge
theory operators $\cO^{l,j,s}$ dual to them
\cite{Ceresole1,Ceresole2}. We have then proceeded in Section
\ref{dgkssec} to work out the two-point functions for these
operators, using the recipe of \cite{dgks,dg}. Just like in the
case of the asymptotically flat three-brane background \cite{gh},
the dilaton equation of motion on the conifold-three-brane
background can be related to Mathieu's equation, which allows an
analytic determination of the corresponding propagator, and
consequently, of the gauge-theory two-point function. Our final
result is given in (\ref{delta2exp}).

Following \cite{dgks}, we have shown that the two-point functions
we obtain are correctly related by the optical theorem, equation
(\ref{optical}), to the absorption probability for the
corresponding dilaton partial wave. For the mode that is symmetric
on $T^{11}$, a further check on our result is that, as seen in
(\ref{delta0lowenergy}), at low energies our two-point function
reduces to the one expected in the context of the Klebanov-Witten
gauge/gravity duality. For higher partial waves, it can be seen in
(\ref{deltallowenergy}) that one also finds a term with the
required Klebanov-Witten behavior, but the leading low-energy
behavior is in fact \emph{singular}. This applies both to the
conifold and the flat-space cases (an important point that was
missed in \cite{dgks}). More work is clearly needed to determine
whether this behavior is real or spurious, and in the former case,
to identify its physical origin.

Equations (\ref{delta0lowenergy}) and (\ref{deltallowenergy}) show
also the leading correction to the two-point functions due to our
departure from the conformal fixed point. As mentioned in Section
\ref{dgkssec}, it would be very interesting to try to test the
mutual consistency of our two main results, (\ref{O8deform}) and
(\ref{delta2exp}), by  showing that this leading correction
matches the pure AdS$_{5}\times T^{11}$ three-point function
$\expec{\cO^{l,j,s}\cO^{l,j,s}\cO_8}$, with $\cO_{8}$ the
dimension-eight operator (\ref{O8}) coming from the deformed
worldvolume Lagrangian (\ref{O8deform}). For the
spherically-symmetric case, the agreement between the two
analogous quantities was demonstrated in \cite{rvr}.

As in the asymptotically flat case, we have seen here, in the
conifold setting, that the recipe of \cite{dgks,dg} yields
correlation functions that are automatically finite, despite the
fact that in (\ref{delta2}) one takes $r$ and $r'$ (which
according to standard AdS/CFT intuition would be associated with a
UV cutoff) to infinity. This is unlike the situation encountered
in applications of the standard GKPW recipe \cite{gkp,w} (which in
some cases can even require momentum-dependent `renormalization',
as in \cite{ms,mr}). An important pending task is to try to
extract from these results information about the precise way in
which the cutoff $\Lambda$ is implemented on the `worldvolume'
side of the duality.

Obvious extensions of our work would be to determine the two-point
functions of operators dual to other supergravity fields, or to
work out higher $n$-point functions. One could also apply the same
methods to explore analogous intermediate-energy dualities in
other interesting backgrounds, like the warped deformed conifold
of \cite{ks}.\footnote{Unfortunately, in this new setting the
dilaton equation of motion is no longer related to Mathieu's
equation, and so one can no longer obtain the complete solution in
analytic form.}

Clearly much work remains to be done to understand the precise
nature of this `worldvolume/geometry' duality at the string theory
level. We would like to return in particular to the subject of the
relation between the duality and the Fischler-Susskind mechanism,
which we brought up in Section \ref{stringsubsec}. Further
progress could perhaps also be made by looking for settings where
the duality can be studied in simplified but still
string-theoretic terms. We believe that a particularly promising
example of this is the Nonrelativistic Wound string theory setting
\cite{km,wound,go}, where the `worldvolume' description of a stack
of (longitudinal) D$p$-branes is provided by NCOS theory
\cite{sst,ncos}, and, as explained in \cite{newt} (see also
\cite{sahakian}),  the `geometry' description of the stack is
afforded by the dual supergravity backgrounds obtained in
\cite{ncos,harmark}.

\section{Acknowledgements}

It is a pleasure to thank Igor Klebanov,  Mart\'{\i}n Kruczenski,
and Juan Maldacena for valuable discussions. We are also grateful
to David Lowe for comments on the manuscript. The work of XA and
HGC has been supported by grant 33951-E from Mexico's National
Council of Science and Technology (CONACyT); that of EC, by a
CONACyT C\'atedra Patrimonial Nivel II; and that of AG, by CONACyT
grants I39233-E and U40745-F.


\begin{thebibliography}{99}


\bibitem{polchrr}
J.~Polchinski, ``Dirichlet-Branes and Ramond-Ramond Charges,''
Phys.\ Rev.\ Lett.\  {\bf 75} (1995) 4724 [arXiv:hep-th/9510017].

\bibitem{dlp}
J.~Dai, R.~G.~Leigh and J.~Polchinski, ``New Connections Between
String Theories,'' Mod.\ Phys.\ Lett.\ A {\bf 4} (1989) 2073.

\bibitem{hs}
G.~T.~Horowitz and A.~Strominger, ``Black Strings And P-Branes,''
Nucl.\ Phys.\ B {\bf 360} (1991) 197.

\bibitem{entropy}
For reviews, see, e.g., \\
J.~M.~Maldacena, ``Black holes in string theory,''
 Princeton University Ph.D. Thesis,
arXiv:hep-th/9607235; \\
G.~T.~Horowitz, ``Quantum states of black holes,''
arXiv:gr-qc/9704072; \\
D.~Youm, ``Black holes and solitons in string theory,'' Phys.\
Rept.\  {\bf 316}, 1 (1999)
[arXiv:hep-th/9710046]; \\
A.~W.~Peet, ``The Bekenstein formula and string theory (N-brane
theory),'' Class.\ Quant.\ Grav.\  {\bf 15}, 3291 (1998)
[arXiv:hep-th/9712253].

\bibitem{3/4}
S.~S.~Gubser, I.~R.~Klebanov and A.~W.~Peet, ``Entropy and
Temperature of Black 3-Branes,'' Phys.\ Rev.\ D {\bf 54} (1996)
3915 [arXiv:hep-th/9602135].

\bibitem{klebabs}
I.~R.~Klebanov, ``World-volume approach to absorption by
non-dilatonic branes,'' Nucl.\ Phys.\ B {\bf 496} (1997) 231
[arXiv:hep-th/9702076].

\bibitem{gkt}
S.~S.~Gubser, I.~R.~Klebanov and A.~A.~Tseytlin, ``String theory
and classical absorption by three-branes,'' Nucl.\ Phys.\ B {\bf
499} (1997) 217 [arXiv:hep-th/9703040].

\bibitem{gk}
S.~S.~Gubser and I.~R.~Klebanov, ``Absorption by branes and
Schwinger terms in the world volume theory,'' Phys.\ Lett.\ B {\bf
413} (1997) 41 [arXiv:hep-th/9708005].

\bibitem{malda}
J.~M.~Maldacena, ``The large N limit of superconformal field
theories and supergravity,'' Adv.\ Theor.\ Math.\ Phys.\  {\bf 2}
(1998) 231 [Int.\ J.\ Theor.\ Phys.\  {\bf 38} (1999) 1113]
[arXiv:hep-th/9711200].

\bibitem{gkp}
S.~S.~Gubser, I.~R.~Klebanov and A.~M.~Polyakov, ``Gauge theory
correlators from non-critical string theory,'' Phys.\ Lett.\ B
{\bf 428} (1998) 105 [arXiv:hep-th/9802109].

\bibitem{w}
E.~Witten, ``Anti-de Sitter space and holography,'' Adv.\ Theor.\
Math.\ Phys.\  {\bf 2} (1998) 253 [arXiv:hep-th/9802150].

\bibitem{magoo}
O.~Aharony, S.~S.~Gubser, J.~M.~Maldacena, H.~Ooguri and Y.~Oz,
``Large N field theories, string theory and gravity,'' Phys.\
Rept.\  {\bf 323} (2000) 183 [arXiv:hep-th/9905111].

\bibitem{ghkk}
S.~S.~Gubser, A.~Hashimoto, I.~R.~Klebanov and M.~Krasnitz,
``Scalar absorption and the breaking of the world volume conformal
invariance,'' Nucl.\ Phys.\ B {\bf 526} (1998) 393
[arXiv:hep-th/9803023].

\bibitem{dealwis}
S.~P.~de Alwis, ``Supergravity, the DBI action and black hole
physics,'' Phys.\ Lett.\ B {\bf 435} (1998) 31
[arXiv:hep-th/9804019].

\bibitem{gh}
S.~S.~Gubser and A.~Hashimoto, ``Exact absorption probabilities
for the D3-brane,'' Commun.\ Math.\ Phys.\  {\bf 203} (1999) 325
[arXiv:hep-th/9805140].

\bibitem{i}
K.~A.~Intriligator, ``Maximally supersymmetric RG flows and AdS
duality,'' Nucl.\ Phys.\ B {\bf 580} (2000) 99
[arXiv:hep-th/9909082].

\bibitem{cm}
N.~R.~Constable and R.~C.~Myers, ``Exotic scalar states in the
AdS/CFT correspondence,'' JHEP {\bf 9911} (1999) 020
[arXiv:hep-th/9905081].

\bibitem{dgks}
U.~H.~Danielsson, A.~G\"uijosa, M.~Kruczenski and B.~Sundborg,
``D3-brane holography,'' JHEP {\bf 0005} (2000) 028
[arXiv:hep-th/0004187].

\bibitem{abks}
O.~Aharony, M.~Berkooz, D.~Kutasov and N.~Seiberg, ``Linear
dilatons, NS5-branes and holography,'' JHEP {\bf 9810} (1998) 004
[arXiv:hep-th/9808149].

\bibitem{ms}
S.~Minwalla and N.~Seiberg, ``Comments on the IIA NS5-brane,''
JHEP {\bf 9906} (1999) 007 [arXiv:hep-th/9904142].

\bibitem{mr}
J.~M.~Maldacena and J.~G.~Russo, ``Large N limit of
non-commutative gauge theories,'' JHEP {\bf 9909} (1999) 025
[arXiv:hep-th/9908134].

\bibitem{dg}
S.~R.~Das and B.~Ghosh, ``A note on supergravity duals of
noncommutative Yang-Mills theory,'' JHEP {\bf 0006} (2000) 043
[arXiv:hep-th/0005007].

\bibitem{hi}
A.~Hashimoto and N.~Itzhaki, ``Non-commutative Yang-Mills and the
AdS/CFT correspondence,'' Phys.\ Lett.\ B {\bf 465} (1999) 142
[arXiv:hep-th/9907166].

\bibitem{rvr}
L.~Rastelli and M.~Van Raamsdonk, ``A note on dilaton absorption
and near-infrared D3 brane holography,'' JHEP {\bf 0012} (2000)
005 [arXiv:hep-th/0011044].

\bibitem{ejp}
N.~Evans, C.~V.~Johnson and M.~Petrini, ``Clearing the throat:
Irrelevant operators and finite temperature in  large N gauge
theory,'' JHEP {\bf 0205} (2002) 002 [arXiv:hep-th/0112058].

\bibitem{hashimoto}
A.~Hashimoto, ``Holographic description of D3-branes in flat
space,'' Phys.\ Rev.\ D {\bf 60} (1999) 127902
[arXiv:hep-th/9903227].

\bibitem{costa}
M.~S.~Costa, ``Absorption by double-centered D3-branes and the
Coulomb branch of N = 4  SYM theory,'' JHEP {\bf 0005} (2000) 041
[arXiv:hep-th/9912073].

\bibitem{costa2}
M.~S.~Costa, ``A test of the AdS/CFT duality on the Coulomb
branch,'' Phys.\ Lett.\ B {\bf 482} (2000) 287 [Erratum-ibid.\ B
{\bf 489} (2000) 439] [arXiv:hep-th/0003289].

\bibitem{dbs}
J.~de Boer and S.~N.~Solodukhin, ``A holographic reduction of
Minkowski space-time,'' arXiv:hep-th/0303006.

\bibitem{co}
P.~Candelas and X.~C.~de la Ossa, ``Comments On Conifolds,''
Nucl.\ Phys.\ B {\bf 342}, 246 (1990).

\bibitem{kw}
I.~R.~Klebanov and E.~Witten, ``Superconformal field theory on
threebranes at a Calabi-Yau  singularity,'' Nucl.\ Phys.\ B {\bf
536} (1998) 199 [arXiv:hep-th/9807080].


\bibitem{dhss}
S.~de Haro, A.~Sinkovics and K.~Skenderis, ``On alpha' corrections
to D-brane solutions,'' arXiv:hep-th/0302136.

\bibitem{gubsereinstein}
S.~S.~Gubser, ``Einstein manifolds and conformal field theories,''
Phys.\ Rev.\ D {\bf 59} (1999) 025006 [arXiv:hep-th/9807164].

\bibitem{kt}
I.~R.~Klebanov and L.~Thorlacius, ``The Size of p-Branes,'' Phys.\
Lett.\ B {\bf 371} (1996) 51 [arXiv:hep-th/9510200].

\bibitem{ghkm}
S.~S.~Gubser, A.~Hashimoto, I.~R.~Klebanov and J.~M.~Maldacena,
``Gravitational lensing by $p$-branes,'' Nucl.\ Phys.\ B {\bf 472}
(1996) 231 [arXiv:hep-th/9601057].

\bibitem{gm}
M.~R.~Garousi and R.~C.~Myers, ``Superstring Scattering from
D-Branes,'' Nucl.\ Phys.\ B {\bf 475} (1996) 193
[arXiv:hep-th/9603194].

\bibitem{bl}
D.~Berenstein and R.~G.~Leigh, ``Superstring perturbation theory
and Ramond-Ramond backgrounds,'' Phys.\ Rev.\ D {\bf 60} (1999)
106002 [arXiv:hep-th/9904104].

\bibitem{dkps}
M.~R.~Douglas, D.~Kabat, P.~Pouliot and S.~H.~Shenker, ``D-branes
and short distances in string theory,'' Nucl.\ Phys.\ B {\bf 485}
(1997) 85 [arXiv:hep-th/9608024].

\bibitem{kv}
J.~Khoury and H.~Verlinde, ``On open/closed string duality,''
Adv.\ Theor.\ Math.\ Phys.\  {\bf 3} (1999) 1893
[arXiv:hep-th/0001056].

\bibitem{polyakov}
A.~M.~Polyakov, ``Gauge fields and space-time,'' Int.\ J.\ Mod.\
Phys.\ A {\bf 17S1} (2002) 119 [arXiv:hep-th/0110196].

\bibitem{dvlmp}
P.~Di Vecchia, A.~Liccardo, R.~Marotta and F.~Pezzella,
``Gauge/gravity correspondence from open/closed string duality,''
arXiv:hep-th/0305061.

\bibitem{acgnr}
T.~Adawi, M.~Cederwall, U.~Gran, B.~E.~Nilsson and B.~Razaznejad,
``Goldstone tensor modes,'' JHEP {\bf 9902} (1999) 001
[arXiv:hep-th/9811145].

\bibitem{hp}
G.~T.~Horowitz and J.~Polchinski, ``A correspondence principle for
black holes and strings,'' Phys.\ Rev.\ D {\bf 55} (1997) 6189
[arXiv:hep-th/9612146].

\bibitem{mt}
R.~R.~Metsaev and A.~A.~Tseytlin, ``Type IIB superstring action in
AdS(5) x S(5) background,'' Nucl.\ Phys.\ B {\bf 533} (1998) 109
[arXiv:hep-th/9805028].

\bibitem{berkovits}
N.~Berkovits, ``Super-Poincare covariant quantization of the
superstring,'' JHEP {\bf 0004} (2000) 018 [arXiv:hep-th/0001035].

\bibitem{bo}
P.~Haggi-Mani and B.~Sundborg, ``Free large N supersymmetric
Yang-Mills theory as a string theory,'' JHEP {\bf 0004} (2000) 031
[arXiv:hep-th/0002189].

\bibitem{dmw}
A.~Dhar, G.~Mandal and S.~R.~Wadia, ``String bits in small radius
AdS and weakly coupled N=4 Super Yang-Mills Theory: I,''
arXiv:hep-th/0304062.

\bibitem{ckky}
A.~Clark, A.~Karch, P.~Kovtun, and D.~Yamada, ``Construction of
bosonic string theory on infinitely curved Anti-de Sitter space,''
arXiv:hep-th/0304107.

\bibitem{v}
H.~Verlinde, ``Bits, matrices and 1/N,'' arXiv:hep-th/0206059.

\bibitem{vv}
D.~Vaman and H.~Verlinde, ``Bit strings from N = 4 gauge theory,''
arXiv:hep-th/0209215.

\bibitem{osft}
For reviews, see, e.g., \\
K.~Ohmori, ``A review on tachyon condensation in open string field
theories,'' arXiv:hep-th/0102085;\\
P.~J.~De Smet, ``Tachyon condensation: Calculations in string
field theory,'' arXiv:hep-th/0109182;\\
W.~Taylor, ``Lectures on D-branes, tachyon condensation, and
string field theory,'' arXiv:hep-th/0301094.

\bibitem{hisft}
A.~Hashimoto and N.~Itzhaki, ``Observables of string field
theory,'' JHEP {\bf 0201} (2002) 028 [arXiv:hep-th/0111092].

\bibitem{ag}
M.~Alishahiha and M.~R.~Garousi, ``Gauge invariant operators and
closed string scattering in open string  field theory,'' Phys.\
Lett.\ B {\bf 536} (2002) 129 [arXiv:hep-th/0201249].

\bibitem{nadav}
N.~Drukker, ``Closed string amplitudes from gauge fixed string
field theory,'' arXiv:hep-th/0207266.

\bibitem{shatash}
S.~L.~Shatashvili, ``On field theory of open strings, tachyon
condensation and closed  strings,'' arXiv:hep-th/0105076.

\bibitem{grsz}
D.~Gaiotto, L.~Rastelli, A.~Sen and B.~Zwiebach, ``Ghost structure
and closed strings in vacuum string field theory,''
arXiv:hep-th/0111129.

\bibitem{gir}
D.~Gaiotto, N.~Itzhaki and L.~Rastelli, ``Closed strings as
imaginary D-branes,'' arXiv:hep-th/0304192.

\bibitem{est}
I.~Ellwood, J.~Shelton and W.~Taylor, ``Tadpoles and Closed String
Backgrounds in Open String Field Theory,'' arXiv:hep-th/0304259.

\bibitem{lovelace}
C.~Lovelace, ``Stability Of String Vacua. 1. A New Picture Of The
Renormalization Group,'' Nucl.\ Phys.\ B {\bf 273} (1986) 413.

\bibitem{fs}
W.~Fischler and L.~Susskind, ``Dilaton Tadpoles, String
Condensates And Scale Invariance,''
Phys.\ Lett.\ B {\bf 171} (1986) 383; \\
``Dilaton Tadpoles, String Condensates And Scale Invariance. 2,''
Phys.\ Lett.\ B {\bf 173} (1986) 262.

\bibitem{dr}
S.~R.~Das and S.~J.~Rey, ``Dilaton Condensates And Loop Effects In
Open And Closed Bosonic Strings,'' Phys.\ Lett.\ B {\bf 186}
(1987) 328.

\bibitem{clny}
C.~G.~Callan, C.~Lovelace, C.~R.~Nappi and S.~A.~Yost, ``String
Loop Corrections To Beta Functions,'' Nucl.\ Phys.\ B {\bf 288}
(1987) 525.

\bibitem{park}
I.~Y.~Park, ``Fundamental vs. solitonic description of D3
branes,'' Phys.\ Lett.\ B {\bf 468} (1999) 213
[arXiv:hep-th/9907142].

\bibitem{pst}
I.~Y.~Park, A.~Sadrzadeh and T.~A.~Tran, ``Super Yang-Mills
operators from the D3-brane action in a curved background,''
Phys.\ Lett.\ B {\bf 497} (2001) 303 [arXiv:hep-th/0010116].

\bibitem{park2}
I.~Y.~Park, ``Strong coupling limit of open strings: Born-Infeld
analysis,'' Phys.\ Rev.\ D {\bf 64} (2001) 081901
[arXiv:hep-th/0106078].

\bibitem{pt}
V.~Periwal and \O.~Tafjord, ``A finite cutoff on the string world
sheet?,'' Phys.\ Rev.\ D {\bf 60} (1999) 046004
[arXiv:hep-th/9803195].

\bibitem{ptrecoil}
V.~Periwal and \O.~Tafjord, ``D-brane recoil,'' Phys.\ Rev.\ D
{\bf 54} (1996) 3690 [arXiv:hep-th/9603156].

\bibitem{fpr}
W.~Fischler, S.~Paban and M.~Rozali, ``Collective Coordinates for
D-branes,'' Phys.\ Lett.\ B {\bf 381} (1996) 62
[arXiv:hep-th/9604014].

\bibitem{cg}
C.~G.~Callan and D.~J.~Gross, ``Quantum Perturbation Theory Of
Solitons,'' Nucl.\ Phys.\ B {\bf 93} (1975) 29.

\bibitem{polchinski}
J.~Polchinski, \emph{String Theory. Vol. 1: An Introduction To The
Bosonic String,} Cambridge University Press (1998), sec.~7.4.

\bibitem{susswi}
L.~Susskind and E.~Witten, ``The holographic bound in anti-de
Sitter space,'' arXiv:hep-th/9805114.

\bibitem{pp}
A.~W.~Peet and J.~Polchinski, ``UV/IR relations in AdS dynamics,''
Phys.\ Rev.\ D {\bf 59} (1999) 065011 [arXiv:hep-th/9809022].

\bibitem{hrg}
See, e.g.,\\
J.~de Boer, ``The holographic renormalization group,'' Fortsch.\
Phys.\  {\bf 49} (2001) 339 [arXiv:hep-th/0101026];\\
K.~Skenderis, ``Lecture notes on holographic renormalization,''
Class.\ Quant.\ Grav.\  {\bf 19} (2002) 5849
[arXiv:hep-th/0209067]; \\
M.~Fukuma, S.~Matsuura and T.~Sakai, ``Holographic renormalization
group,'' Prog.\ Theor.\ Phys.\  {\bf 109} (2003) 489
[arXiv:hep-th/0212314]\\
and references therein.


\bibitem{mp}
D.~Morrison and R.~Plesser, ``Non-spherical horizons: I,'' Adv.\
Theor.\ Math.\ Phys.\ {\bf 3} (1999) 1 [arXiv:hep-th/9810201].

\bibitem{hko}
C.~P.~Herzog, I.~R.~Klebanov and P.~Ouyang, ``D-branes on the
conifold and N=1 gauge/gravity dualities,'' arXiv:hep-th/0205100.

\bibitem{afhs}
B.~S.~Acharya, J.~M.~Figueroa-O'Farrill, C.~M.~Hull and B.~Spence,
``Branes at conical singularities and holography,'' Adv.\ Theor.\
Math.\ Phys.\  {\bf 2} (1999) 1249 [arXiv:hep-th/9808014].

\bibitem{Ceresole1}
A.~Ceresole, G.~Dall'Agata, R.~D'Auria and S.~Ferrara, ``Spectrum
of type IIB supergravity on AdS(5) x T(11): Predictions on N  = 1
SCFT's,'' Phys.\ Rev.\ D {\bf 61} (2000) 066001
[arXiv:hep-th/9905226].

\bibitem{Ceresole2}
A.~Ceresole, G.~Dall'Agata and  R.~D'Auria, ``KK Spectroscopy  of
Type IIB Supergravity on AdS(5) x T(11),'' JHEP {\bf 9911}, 009,
(1999); [arXiv:hep-th/9907216].

\bibitem{jrd}
D.~P.~Jatkar and S.~Randjbar-Daemi, ``Type IIB string theory on
AdS(5) x T(n n'),'' Phys.\ Lett.\ B {\bf 460} (1999) 281
[arXiv:hep-th/9904187].

\bibitem{abs}
O.~Aharony, M.~Berkooz and E.~Silverstein, ``Multiple-trace
operators and non-local string theories,'' JHEP {\bf 0108} (2001)
006
[arXiv:hep-th/0105309]; \\
``Non-local string theories on AdS(3) x S**3 and stable
non-supersymmetric backgrounds,'' Phys.\ Rev.\ D {\bf 65} (2002)
106007 [arXiv:hep-th/0112178].

\bibitem{wittenmultitr}
E.~Witten, ``Multi-trace operators, boundary conditions and
AdS/CFT correspondence,'' [arxiv:hep-th/0112258].

\bibitem{bss}
M.~Berkooz, A.~Sever and A.~Shomer, ``Double-trace deformations,
boundary conditions and spacetime  singularities,'' JHEP {\bf
0205} (2002) 034 [arXiv:hep-th/0112264].

\bibitem{ss}
A.~Sever and A.~Shomer, ``A note on multi-trace deformations and
AdS/CFT,'' JHEP {\bf 0207} (2002) 027 [arXiv:hep-th/0203168].


\bibitem{KTR}
I.~R.~Klebanov, W.~I.~Taylor and M.~Van Raamsdonk, ``Absorption of
dilaton partial waves by D3-branes,'' Nucl.\ Phys.\ B {\bf 560}
(1999) 207 [arXiv:hep-th/9905174].

\bibitem{LMR}
S.~M.~Lee, S.~Minwalla, M.~Rangamani and N.~Seiberg, ``Three-point
functions of chiral operators in D = 4, N = 4 SYM at large N,''
Adv.\ Theor.\ Math.\ Phys.\  {\bf 2} (1998) 697
[arXiv:hep-th/9806074].

\bibitem{ks2}
I.~R.~Klebanov, M.~J.~Strassler ``Supergravity and a Confining
Gauge Theory: Duality Cascades and $\chi$SB-Resolution of Naked
Singularities,'' JHEP {\bf 0008} (2000) 052 \
[arXiv:hep-th/0007191].


\bibitem{mmlz}
R.~Manvelyan, H.~J.~M\"uller-Kirsten, J.~Q.~Liang and Y.~Zhang,
``Absorption cross section of scalar field in supergravity
background,'' Nucl.\ Phys.\ B {\bf 579} (2000) 177
[arXiv:hep-th/0001179].

\bibitem{clvp}
M.~Cveti\v{c}, H.~L\"u and J.~F.~V\'azquez-Poritz, ``Absorption by
extremal D3-branes,'' JHEP {\bf 0102} (2001) 012
[arXiv:hep-th/0002128].

\bibitem{mvrs}
S.~Minwalla, M.~Van Raamsdonk and N.~Seiberg, ``Noncommutative
perturbative dynamics,'' JHEP {\bf 0002} (2000) 020
[arXiv:hep-th/9912072].

\bibitem{vrs}
M.~Van Raamsdonk and N.~Seiberg, ``Comments on noncommutative
perturbative dynamics,'' JHEP {\bf 0003} (2000) 035
[arXiv:hep-th/0002186].


\bibitem{drnc}
S.~R.~Das and S.~J.~Rey, ``Open Wilson lines in noncommutative
gauge theory and tomography of  holographic dual supergravity,''
Nucl.\ Phys.\ B {\bf 590} (2000) 453 [arXiv:hep-th/0008042].

\bibitem{ks}
I.~R.~Klebanov and M.~J.~Strassler, ``Supergravity and a confining
gauge theory: Duality cascades and chiSB-resolution of naked
singularities,'' JHEP {\bf 0008} (2000) 052
[arXiv:hep-th/0007191].

\bibitem{km}
I.~R.~Klebanov and J.~M.~Maldacena, ``1+1 dimensional NCOS and its
U(N) gauge theory dual,'' Int.\ J.\ Mod.\ Phys.\ A {\bf 16} (2001)
922 [Adv.\ Theor.\ Math.\ Phys.\  {\bf 4} (2000) 283]
[arXiv:hep-th/0006085].

\bibitem{wound}
U.~H.~Danielsson, A.~G\"uijosa and M.~Kruczenski, ``IIA/B, wound
and wrapped,'' JHEP {\bf 0010} (2000) 020 [arXiv:hep-th/0009182].

\bibitem{go}
J.~Gomis and H.~Ooguri, ``Non-relativistic closed string theory,''
J.\ Math.\ Phys.\  {\bf 42} (2001) 3127 [arXiv:hep-th/0009181].

\bibitem{sst}
N.~Seiberg, L.~Susskind and N.~Toumbas, ``Strings in background
electric field, space/time noncommutativity and a new noncritical
string theory,'' JHEP {\bf 0006} (2000) 021
[arXiv:hep-th/0005040].

\bibitem{ncos}
R.~Gopakumar, J.~M.~Maldacena, S.~Minwalla and A.~Strominger,
``S-duality and noncommutative gauge theory,'' JHEP {\bf 0006}
(2000) 036 [arXiv:hep-th/0005048].

\bibitem{newt}
U.~H.~Danielsson, A.~G\"uijosa and M.~Kruczenski, ``Newtonian
gravitons and D-brane collective coordinates in wound string
theory,'' JHEP {\bf 0103} (2001) 041 [arXiv:hep-th/0012183].

\bibitem{sahakian}
V.~Sahakian, ``The large M limit of non-commutative open strings
at strong coupling,'' Nucl.\ Phys.\ B {\bf 621} (2002) 62
[arXiv:hep-th/0107180].

\bibitem{harmark}
T.~Harmark, ``Supergravity and space-time non-commutative open
string theory,'' JHEP {\bf 0007} (2000) 043
[arXiv:hep-th/0006023].

\end{thebibliography}
\end{document}